\documentclass[12pt]{article}
\usepackage[english]{babel}
\newcommand{\1}{1 \!\! 1}

\usepackage{tikz}
\usepackage{amsmath,amsthm,amssymb,amsfonts}
\usepackage{color}
\usepackage{amsfonts}
\usepackage{graphicx}
\usepackage{float}
\usepackage{cite}

\usepackage[colorlinks=true,linkcolor=blue,citecolor=red, urlcolor=green]{hyperref} % per colorare citazioni, link, email ecc

\usepackage[labelformat=simple]{subfig}

\begin{document}

\title{Spreading of information on a network:\\ a quantum view}
\author{F.~Bagarello$^{1,2}$,
F. Gargano$^{1}$,
M.~Gorgone$^{3}$ and 
F.~Oliveri$^{3}$
\ \\\\
{\footnotesize $^{1}$ Dipartimento di Ingegneria -  Universit\`a di Palermo,}\\
{\footnotesize Viale delle Scienze, I--90128  Palermo, Italy;}\\ {\footnotesize fabio.bagarello@unipa.it; francesco.gargano@unipa.it}\\
{\footnotesize $^{2}$ I.N.F.N -  Sezione di Catania, Italy}\\
{\footnotesize $^{3}$ Dipartimento di Scienze Matematiche e Informatiche, Scienze Fisiche e Scienze della Terra,}\\
{\footnotesize
Universit\`a di Messina, Viale F. Stagno d'Alcontres 31, 98166, Messina, Italy;}\\
{\footnotesize matteo.gorgone@unime.it; francesco.oliveri@unime.it}
}

\date{Published in \textit{Entropy} \textbf{25}, 1438 (2023).}

\maketitle

\begin{abstract}
This paper concerns the modeling of the spread of information through a complex, multi-layered network, where the information is transferred from an initial transmitter to a final receiver. The mathematical model is deduced within the framework of operatorial methods, 
according to the formal mathematical apparatus typical of quantum 
mechanics. Two different approaches are considered: one based on
the ($H,\rho$)-induced dynamics and one on the Gorini--Kossakowski--Sudarshan--Lindblad (GKSL) equation. For each method, numerical results are presented.
\end{abstract}


\section{Introduction}
The rapid development of social media and collaborative web platforms had drastically altered how information is disseminated, making users both creators and transmitters of news. Although this global evolution had accelerated the spread of information, it had introduced significant challenges in determining the information's reliability. In fact, information can be readily manipulated, skewed, or partially omitted, resulting in widespread misinformation, baseless rumours, or outright falsehoods.
Consequently, it is not surprising that there has been an overall rise in the development of mathematical models designed to capture these phenomena. Many models are based on the framework usually adopted in epidemiological models and graph analyses  \cite{N1, N2, N3, N4}, but also operatorial methods \cite{fff1}
have been used.

In this paper, we continue the analysis first undertaken in \cite{fff1} with the goal of modeling the spread of information through  a complex, multi-layered network, where the information is transferred from an initial transmitter to a final receiver. The intermediate layers in this model, besides transferring the information, may possibly distort it. 
Our approach is based on the idea of exchanging packets of information and uses some of the mathematical apparatuses of quantum mechanics; the main motivation for this specific choice is that, in the last few years, operatorial techniques have been used to model successfully several macroscopic systems (see the recent monographs \cite{bookBGO,Rob2023} and references therein).
In particular, and to cover a large spectrum of possibilities, we shall adopt two different approaches: one based on the $(H,\rho)$-induced dynamics \cite{HRO2} and one on the  Gorini--Kossakowski--Sudarshan--Lindblad (GKSL) equation \cite{Manz}. More specifically, concerning the $(H,\rho)$-induced dynamics approach, our framework involves a Hamiltonian operator $H$ built with appropriate ladder operators, followed by inferring the Heisenberg equations of motion to derive the time evolution of the significant observables of the system. According to the $(H,\rho)$-induced dynamics methodology \cite{bookBGO,HRO2}, we introduce a 
rule-oriented dynamics typical of cellular automata into the Heisenberg dynamics. This approach enables us to incorporate into our model certain effects that cannot easily be described by a purely Hamiltonian approach. 
In the second part of the paper, when dealing with the GKSL equation, we shall mainly obtain the dynamics by introducing suitable Lindblad operators \cite{Manz} and compare these results with those deduced with the previous approach. 
A quantitative measure of the reliability of the information transmitted is achieved through the computation of the mean values of suitable density operators. One key advantage of this second approach is its ability to naturally implement the irreversible nature of the transmission mechanisms between the various layers, reflecting the fact that (usually) transmitters do not receive information from the lower receiving layers.

The aim of both models is to assess, within the inherent uncertainty characterizing the spread of information via unreliable agents, whether a final receiver interprets information as good or fake. This  is achieved through the computation of certain mean values, which are phenomenologically interpreted as a measure of the information credibility or lack thereof. We would like to emphasize that the terms 'good' and 'fake' are not limited strictly to their usual definitions, but can represent any dichotomy within the information. For example, a piece of news can be conveyed emphasizing the benefits (interpreted as 'good') or drawbacks (equated to 'fake') of a particular subject. 
The ultimate goal is to characterize the innate uncertainty linked to the propagation of information through unreliable channels, primarily social media. We use the models in some scenarios that incorporate unique behaviors of the agents (i.e., different rules $\rho$ or different rates of transmission between the layers) that are able to significantly transform how the information is perceived by the receiver.

The paper is structured as follows. In Section~\ref{sec::sec2}, we present the macroscopic system and the mathematical model built according to the Heisenberg dynamics of the operators. In Section~\ref{sec:model}, we present some numerical experiments based on the $(H,\rho)$-induced dynamics with different rules for a multi-layered network, whereas Section~\ref{section::GKSL} is devoted to the description of the dynamics of the network with a model derived from the GKSL equation and suitable Lindblad operators.
Finally, Section \ref{section:Concl} summarizes our conclusions.

\section{The Model and Its Dynamics}
\label{sec::sec2}
The mathematical framework and the Hamiltonian of the model were already introduced in \cite{fff1}. To keep this paper almost self-contained, hereafter, we give the essential steps of our construction.

We have a piece of news $\mathcal{N}$ that can be transmitted as it is or changed by the agent who transmits $\mathcal{N}$. In the first case, we speak of \emph{{good news}}, 
 while we define {\em {fake news}} when  a piece of news is somehow distorted, for the agent's convenience, ignorance, or other reasons.  We suppose that we have $N$ agents $A_\alpha$, with $\alpha=1,2,\ldots,N$, creating, receiving, and transmitting $\mathcal{N}$. Each agent is seen as a different cell of a network $\mathcal{R}$. Two cells $\alpha$ and $\beta$ are neighboring if the agents $A_\alpha$ and  $A_\beta$ have a  direct link to interchange information, while, in most cases, $A_\alpha$ is connected to $A_\beta$ only by means of intermediate agents. As in \cite{fff1}, for each $A_\alpha$, we introduce two families of fermionic operators, $f_\alpha$ {(for fake news)} and $g_\alpha$ {(for good news)}, satisfying the canonical anticommutation rules (CAR)
\begin{equation}
\{f_\alpha,f_\beta^\dagger\}=\{g_\alpha,g_\beta^\dagger\}=\delta_{\alpha,\beta}\1, 
\label{21}
\end{equation}
where $\{x,y\}=xy+yx$, with all the other anticommutators being trivial. In particular, $\{f_\alpha,f_\beta\}=\{g_\alpha,g_\beta\}=0$, and 
$\{f_\alpha^\sharp,g_\beta^\sharp\}=0$, where $x^\sharp=x$ or $x^\sharp=x^\dagger$. We further define the number operators $\widehat{F}_\alpha=f_\alpha^\dagger f_\alpha$ and $\widehat{G}_\alpha=g_\alpha^\dagger g_\alpha$ and the four-dimensional Hilbert space $\mathcal{H}_\alpha$ as follows: we introduce first the vacua $e_{\alpha,0}^{(f)}$ and $e_{\alpha,0}^{(g)}$ of $f_\alpha$
and $g_\alpha$, respectively: $f_\alpha\,e_{\alpha,0}^{(f)}=g_\alpha\,e_{\alpha,0}^{(g)}=0$. Then, we define 
\[
e_{\alpha,1}^{(f)}=f_\alpha^\dagger e_{\alpha,0}^{(f)}, \qquad e_{\alpha,1}^{(g)}=g_\alpha^\dagger e_{\alpha,0}^{(g)},
\]
and
\begin{equation}
\varphi_{\alpha:n_f,n_g}=e_{\alpha,n_f}^{(f)} \otimes e_{\alpha,n_g}^{(g)},
\label{22}
\end{equation}
where $n_f,n_g=0,1$. The set $\mathcal{F}_\varphi(\alpha)=\{\varphi_{\alpha:n_f,n_g}\}$ is an orthonormal (o.n.) basis of $\mathcal{H}_\alpha$. We call $\langle.,.\rangle_\alpha$ the scalar product in $\mathcal{H}_\alpha$, and we have
$$
\langle\varphi_{\alpha:n_f,n_g},\varphi_{\alpha:m_f,m_g}\rangle_\alpha=\delta_{n_f,m_f}\delta_{n_g,m_g}.
$$

Moreover,
\begin{equation}
\widehat{F}_\alpha \varphi_{\alpha:n_f,n_g}=n_f\,\varphi_{\alpha:n_f,n_g}, \qquad \widehat{G}_\alpha \varphi_{\alpha:n_f,n_g}=n_g\,\varphi_{\alpha:n_f,n_g}.
\label{23}
\end{equation}
As already proposed in \cite{fff1}, if the system $\mathcal{S}$ in $\alpha$ is described by the vector 
$\varphi_{\alpha:0,0}$, then $\mathcal{N}$ has not reached $A_\alpha$, in any of its forms. If it is described by  $\varphi_{\alpha:1,0}$, then the fake version of $\mathcal{N}$ has reached $A_\alpha$, while $A_\alpha$ is reached by its good version if the vector is $\varphi_{\alpha:0,1}$. Finally, both versions of $\mathcal{N}$ have reached $A_\alpha$ if $\mathcal{S}$ in $\alpha$ is described by $\varphi_{\alpha:1,1}$.

Each vector $u_\alpha\in\mathcal{H}_\alpha$ is a linear combination of the vectors $\varphi_{\alpha:n_f,n_g}$. Now, we can consider $\mathcal{H}=\otimes_\alpha \mathcal{H}_\alpha$, the Hilbert space of $\mathcal{R}$, with scalar product
$$
\langle u,v\rangle=\prod_\alpha \langle u_\alpha,v_\alpha\rangle_\alpha,
$$
for each $u=\otimes_\alpha u_\alpha$ and $v=\otimes_\alpha v_\alpha$. Each operator $\widehat{X}_\alpha$ acting on $\mathcal{H}_\alpha$ can be extended to the whole $\mathcal{H}$ by identifying $\widehat{X}_\alpha$ with $\widehat{X}_\alpha\otimes(\otimes_{\beta\neq\alpha} \1_\beta)$, where $\1_\beta$ is the identity operator on $\mathcal{H}_\beta$. The initial state of $\mathcal{S}$ is described by the following vector on $\mathcal{H}$:
\begin{equation}
\Psi_{\bf n, m}=\otimes_\alpha \varphi_{\alpha:n_\alpha,m_\alpha},
\label{24}
\end{equation}
where ${\bf n}=(n_1,n_2,\ldots,n_N)$, ${\bf m}=(m_1,m_2,\ldots,m_N)$. The knowledge of $\Psi_{\bf n, m}$ allows us to deduce whether and which kind of information is possessed (at $t=0$) by any agent $A_\alpha$ of the~system.

To assign a dynamics to the system, let us introduce the following Hamiltonian $H$ of the system, describing the main interactions occurring in $\mathcal{S}$ \cite{fff1}:
\begin{equation}
\left\{
\begin{aligned}
H&=H_0+ H_I, \qquad \hbox{ with }  \\
H_0 &=\sum_{\alpha} \omega_{f,\alpha}\widehat{F}_\alpha+\sum_{\alpha}\omega_{g,\alpha}\widehat{G}_\alpha,\\
H_I &= \sum_{\alpha,\beta}\,p_{\alpha,\beta}^{(f)}(f_\alpha f_\beta^\dagger+f_\beta f_\alpha^\dagger)+\sum_{\alpha,\beta}\,p_{\alpha,\beta}^{(g)}(g_\alpha g_\beta^\dagger+g_\beta g_\alpha^\dagger)+\sum_{\alpha}\,\lambda_{\alpha}(f_\alpha g_\alpha^\dagger+g_\alpha f_\alpha^\dagger).
\end{aligned}
\right.
\label{25}
\end{equation}
We refer to \cite{fff1} for some details of this Hamiltonian. Here, we limit ourselves to observing that the contribution $H_0$ is the free part of the Hamiltonian, the parameters therein involved being somehow related to the \emph{{inertia}} of the corresponding degrees of freedom. Moreover, the first term in $H_I$ is a diffusion term for fake news, while the second one is again a diffusion term, but for good news. The third term describes a possible  change during the time evolution of the nature of the news: a piece of good news can be transformed into a fake one ($g_\alpha f_\alpha^\dagger$) and vice versa (by means of the adjoint contribution). The value of $\lambda_\alpha$ measures the attitude of $A_\alpha$ to modify the news.  The coefficients $p_{\alpha,\beta}^{(f,g)}$ in $H_I$ are \emph{{diffusion coefficients}} for the two typologies of news. We assume for the moment that they are all real and symmetric ($p_{\alpha,\beta}^{(f,g)}=p_{\beta,\alpha}^{(f,g)}$), and that $p_{\alpha,\alpha}^{(f,g)}=0$, but they are not necessarily all different from zero. 

In \cite{fff1}, we worked under the assumption that
$$
\sum_{\alpha,\beta}p_{\alpha,\beta}^{(f)}>\sum_{\alpha,\beta}p_{\alpha,\beta}^{(g)},
$$
to describe the fact  that fake news travels much faster than good news. Hereafter, we do not necessarily adopt this assumption.

{Using standard computation, we use $H$ and the CAR in Equation \eqref{21} to compute} the equations of motion for the ladder operators and the mean values of their related number operators. In particular, using $\dot X(t)=i[H,X(t)]$, where $[x,y]=xy-yx$,
we obtain
\begin{equation}
\left\{
\begin{array}{ll}
\dot f_\alpha(t)=-i\omega_{f,\alpha}\,f_\alpha(t)+2i\sum_\beta p_{\alpha,\beta}^{(f)}f_\beta(t)+i\lambda_\alpha g_\alpha(t)  \\
\dot g_\alpha(t)=-i\omega_{g,\alpha}\,g_\alpha(t)+2i\sum_\beta p_{\alpha,\beta}^{(g)}g_\beta(t)+i\lambda_\alpha f_\alpha(t), 
\end{array}
\right.
\label{26}
\end{equation}
where $\alpha=1,2,\ldots,N$. This is a closed system of linear, operator-valued, first-order differential equations that can be easily solved. In fact, if we call $X(t)$ the $2N$-column vectors whose transpose is
$$
X(t)^T=\left(f_1(t),f_2(t),\ldots,f_N(t),g_1(t),g_2(t),\ldots,g_N(t)\right),
$$ 
and introduce the following Hermitian (time-independent) $2N\times2N$ matrix $V$,
$$
V=\left(
\begin{array}{cccccccccccc}
-\omega_{f,1} & 2p_{1,2}^{(f)} &  2p_{1,3}^{(f)} & . & . &  2p_{1,N}^{(f)} & \lambda_1 & 0 & 0 & . & . & 0\\
 2p_{1,2}^{(f)} & -\omega_{f,2} &  2p_{2,3}^{(f)} & . & . &  2p_{2,N}^{(f)} & 0 & \lambda_2  & 0 & . & . & 0\\
2p_{1,3}^{(f)} &  2p_{3,2}^{(f)} & -\omega_{f,3} & . & . &  2p_{3,N}^{(f)} & 0 & 0 & \lambda_3 & . & . & 0\\
. & . & . & . & . &  . & . & . & . & . & . & .\\
. & . & . & . & . &  . & . & . & . & . & . & .\\
2p_{1,N}^{(f)} &  2p_{2,N}^{(f)} & 2p_{3,N}^{(f)} & . & . &   -\omega_{f,N} & 0 & 0 & . & . & . & \lambda_N\\
\lambda_1 & 0 & 0 & . & . & 0 & -\omega_{g,1} & 2p_{1,2}^{(g)} &  2p_{1,3}^{(g)} & . & . &  2p_{1,N}^{(g)} \\
0 & \lambda_2  & 0 & . & . & 0 &  2p_{1,2}^{(g)} & -\omega_{g,2} &  2p_{2,3}^{(g)} & . & . &  2p_{2,N}^{(g)}\\
0 & 0 & \lambda_3 & . & . & 0 & 2p_{1,3}^{(g)} &  2p_{3,2}^{(g)} & -\omega_{g,3} & . & . &  2p_{3,N}^{(g)}\\
. & . & . & . & . &  . & . & . & . & . & . & .\\
. & . & . & . & . &  . & . & . & . & . & . & .\\
0 & 0 & . & . & . & \lambda_N & 2p_{1,N}^{(g)} &  2p_{2,N}^{(g)} & 2p_{3,N}^{(g)} & . & . &   -\omega_{g,N} \\
\end{array}
\right)
$$
system \eqref{26} can be rewritten in compact form as $\dot X(t)=iVX(t)$, whereupon the solution is
\begin{equation}
X(t)=\exp(iVt)X(0).
\label{27}
\end{equation}
In writing the explicit form of $V$, we have used the equality $p_{\alpha,\beta}^{(f,g)}=p_{\beta,\alpha}^{(f,g)}$. Let us call $v_{i,j}(t)=\left(\exp(iVt)\right)_{i,j}$ $(i,j=1,2,\ldots,2N)$, and let $\mathcal{E}=\{e_j, \; j=1,2,\ldots,2N\}$ be the canonical o.n. basis in $\mathcal{H}_{2N}=\mathbb{C}^{2N}$, endowed with scalar product $\langle\cdot,\cdot\rangle_{2N}$. Then, we~have
\begin{equation}
f_\alpha(t)=\langle e_\alpha, X(t)\rangle_{2N}, \qquad g_\alpha(t)=\langle e_{N+\alpha}, X(t)\rangle_{2N},
\label{28}
\end{equation}
$\alpha=1,2,\ldots,N$. If we call $F_\beta^0$ and $G_\beta^0$ the mean value of $\widehat{F}_\beta$ and $\widehat{G}_\beta$ on the vector $\varphi_{\beta:n_\beta,m_\beta}$ at $t=0$,
$$
F_\beta^0=\langle\varphi_{\beta:n_\beta,m_\beta}, \widehat{F}_\beta\varphi_{\beta:n_\beta,m_\beta}\rangle_\beta, \qquad G_\beta^0=\langle\varphi_{\beta:n_\beta,m_\beta}, \widehat{G}_\beta\varphi_{\beta:n_\beta,m_\beta}\rangle_\beta,
$$
then, setting
\begin{equation}
F_\alpha(t)=\langle\Psi_{\bf n, m}, \widehat{F}_\alpha(t)\Psi_{\bf n, m}\rangle_{2N}=\langle\Psi_{\bf n, m},  f_\alpha^\dagger(t)f_\alpha(t)\Psi_{\bf n, m}\rangle_{2N},
\label{29}
\end{equation} 
and
\begin{equation}
G_\alpha(t)=\langle\Psi_{\bf n, m}, \widehat{G}_\alpha(t)\Psi_{\bf n, m}\rangle_{2N}=\langle\Psi_{\bf n, m},  g_\alpha^\dagger(t)g_\alpha(t)\Psi_{\bf n, m}\rangle_{2N},
\label{210}
\end{equation}
we obtain
\begin{equation}
\left\{
\begin{aligned}
&F_\alpha(t)=\sum_{\beta=1}^N\left(|v_{\alpha,\beta}(t)|^2F_\beta^0+|v_{\alpha,\beta+N}(t)|^2G_\beta^0\right),
\\ 
&G_\alpha(t)=\sum_{\beta=1}^N\left(|v_{\alpha+N,\beta}(t)|^2F_\beta^0+|v_{\alpha+N,\beta+N}(t)|^2G_\beta^0\right), 
\end{aligned}
\right.
\label{211}
\end{equation}
$\alpha=1,2,\ldots,N$. From these functions, we define the following mean values
\begin{equation}
F(t)=\frac{1}{N}\,\sum_{\alpha=1}^NF_\alpha(t), \qquad G(t)=\frac{1}{N}\,\sum_{\alpha=1}^NG_\alpha(t),
\label{212}
\end{equation}
which we interpret as the time evolution of the  \emph{global} mean values of fake and good news,
respectively. On the other hand, $F_\alpha(t)$ and $G_\alpha(t)$ are their \emph{local} counterparts. Notice that, because of the fermionic nature of the operators involved, we have $F_\alpha(t), G_\alpha(t)\in[0,1]$, and therefore  $F(t), G(t)\in[0,1]$, as well, for all values of $t$.

The resulting time evolution is, in general, quasiperiodic, so that neverending oscillations occur. This is a general fact (see \cite{bagbook1,bagbook2}), due to the fact that our Hamiltonian in Equation (\ref{25}) is Hermitian and that $\mathcal{S}$ is finite-dimensional.
As we will show below,  some more interesting dynamical behaviors can be recovered when introducing some more components in the analysis of $\mathcal{S}$, such as using the $(H,\rho)$-induced dynamics approach or Lindblad~operators. 
   
\section{The Network Model}
\label{sec:model}

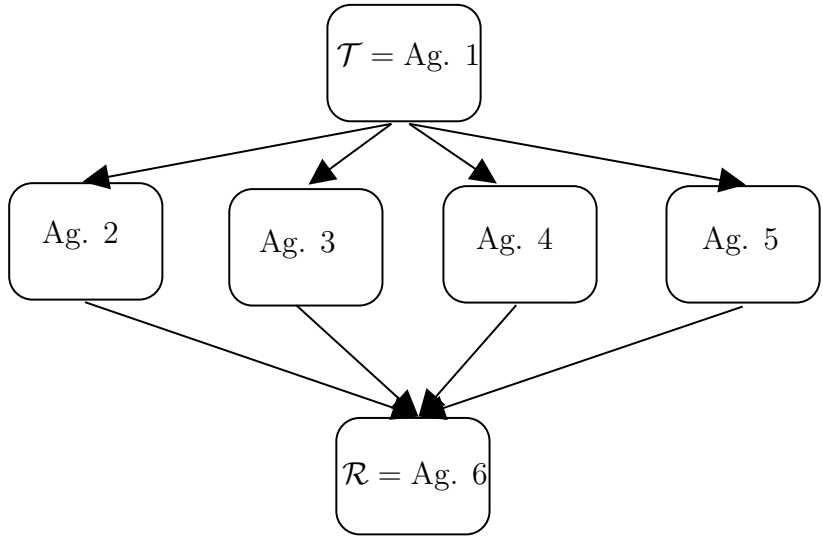
\begin{figure}[H] 
	\tikzset{every picture/.style={line width=0.75pt}} %set default line width to 0.75pt        
	\hspace*{1.25cm}\begin{tikzpicture}[x=0.75pt,y=0.75pt,yscale=-1,xscale=1,scale=1.5]
		%uncomment if require: \path (0,266); %set diagram left start at 0, and has height of 266
		
		%Rounded Rect [id:dp561861258256726] 
		\draw   (187,109.84) .. controls (187,105.51) and (190.51,102) .. (194.84,102) -- (230.68,102) .. controls (235.01,102) and (238.51,105.51) .. (238.51,109.84) -- (238.51,133.34) .. controls (238.51,137.67) and (235.01,141.18) .. (230.68,141.18) -- (194.84,141.18) .. controls (190.51,141.18) and (187,137.67) .. (187,133.34) -- cycle ;
		%Rounded Rect [id:dp5085602796495265] 
		\draw   (261,111.84) .. controls (261,107.51) and (264.51,104) .. (268.84,104) -- (304.68,104) .. controls (309.01,104) and (312.51,107.51) .. (312.51,111.84) -- (312.51,135.34) .. controls (312.51,139.67) and (309.01,143.18) .. (304.68,143.18) -- (268.84,143.18) .. controls (264.51,143.18) and (261,139.67) .. (261,135.34) -- cycle ;
		%Rounded Rect [id:dp1723621049764612] 
		\draw   (297,188.84) .. controls (297,184.51) and (300.51,181) .. (304.84,181) -- (340.68,181) .. controls (345.01,181) and (348.51,184.51) .. (348.51,188.84) -- (348.51,212.34) .. controls (348.51,216.67) and (345.01,220.18) .. (340.68,220.18) -- (304.84,220.18) .. controls (300.51,220.18) and (297,216.67) .. (297,212.34) -- cycle ;
		%Straight Lines [id:da29615257959469843] 
		\draw    (321.8,179.53) -- (212.5,142.04) ;
		\draw [shift={(324.64,180.51)}, rotate = 198.93] [fill={rgb, 255:red, 0; green, 0; blue, 0 }  ][line width=0.08]  [draw opacity=0] (8.93,-4.29) -- (0,0) -- (8.93,4.29) -- cycle    ;
		%Rounded Rect [id:dp3426002489793383] 
		\draw   (408,110.84) .. controls (408,106.51) and (411.51,103) .. (415.84,103) -- (451.68,103) .. controls (456.01,103) and (459.51,106.51) .. (459.51,110.84) -- (459.51,134.34) .. controls (459.51,138.67) and (456.01,142.18) .. (451.68,142.18) -- (415.84,142.18) .. controls (411.51,142.18) and (408,138.67) .. (408,134.34) -- cycle ;
		%Rounded Rect [id:dp6888081637183974] 
		\draw   (333,110.84) .. controls (333,106.51) and (336.51,103) .. (340.84,103) -- (376.68,103) .. controls (381.01,103) and (384.51,106.51) .. (384.51,110.84) -- (384.51,134.34) .. controls (384.51,138.67) and (381.01,142.18) .. (376.68,142.18) -- (340.84,142.18) .. controls (336.51,142.18) and (333,138.67) .. (333,134.34) -- cycle ;
		%Straight Lines [id:da7587438998096208] 
		\draw    (283.5,143.04) -- (322.42,178.49) ;
		\draw [shift={(324.64,180.51)}, rotate = 222.33] [fill={rgb, 255:red, 0; green, 0; blue, 0 }  ][line width=0.08]  [draw opacity=0] (8.93,-4.29) -- (0,0) -- (8.93,4.29) -- cycle    ;
		%Straight Lines [id:da010749630937346932] 
		\draw    (321.51,82.03) -- (431.69,101.97) ;
		\draw [shift={(434.64,102.51)}, rotate = 190.26] [fill={rgb, 255:red, 0; green, 0; blue, 0 }  ][line width=0.08]  [draw opacity=0] (8.93,-4.29) -- (0,0) -- (8.93,4.29) -- cycle    ;
		%Straight Lines [id:da8209874717598136] 
		\draw    (315.51,82.03) -- (214.59,100.95) ;
		\draw [shift={(211.64,101.51)}, rotate = 349.38] [fill={rgb, 255:red, 0; green, 0; blue, 0 }  ][line width=0.08]  [draw opacity=0] (8.93,-4.29) -- (0,0) -- (8.93,4.29) -- cycle    ;
		%Rounded Rect [id:dp6231508624823057] 
		\draw   (294,49.84) .. controls (294,45.51) and (297.51,42) .. (301.84,42) -- (337.68,42) .. controls (342.01,42) and (345.51,45.51) .. (345.51,49.84) -- (345.51,73.34) .. controls (345.51,77.67) and (342.01,81.18) .. (337.68,81.18) -- (301.84,81.18) .. controls (297.51,81.18) and (294,77.67) .. (294,73.34) -- cycle ;
		%Straight Lines [id:da16497902925557995] 
		\draw    (321.51,82.03) -- (348.52,100.46) ;
		\draw [shift={(351,102.15)}, rotate = 214.3] [fill={rgb, 255:red, 0; green, 0; blue, 0 }  ][line width=0.08]  [draw opacity=0] (8.93,-4.29) -- (0,0) -- (8.93,4.29) -- cycle    ;
		%Straight Lines [id:da44590542074775663] 
		\draw    (315.51,82.03) -- (290.06,100.73) ;
		\draw [shift={(287.64,102.51)}, rotate = 323.7] [fill={rgb, 255:red, 0; green, 0; blue, 0 }  ][line width=0.08]  [draw opacity=0] (8.93,-4.29) -- (0,0) -- (8.93,4.29) -- cycle    ;
		%Straight Lines [id:da07154436775879325] 
		\draw    (357.5,143.04) -- (326.62,178.25) ;
		\draw [shift={(324.64,180.51)}, rotate = 311.26] [fill={rgb, 255:red, 0; green, 0; blue, 0 }  ][line width=0.08]  [draw opacity=0] (8.93,-4.29) -- (0,0) -- (8.93,4.29) -- cycle    ;
		%Straight Lines [id:da8239777787049356] 
		\draw    (433.64,143.51) -- (327.48,179.54) ;
		\draw [shift={(324.64,180.51)}, rotate = 341.25] [fill={rgb, 255:red, 0; green, 0; blue, 0 }  ][line width=0.08]  [draw opacity=0] (8.93,-4.29) -- (0,0) -- (8.93,4.29) -- cycle    ;
		
		% Text Node
		\draw (197,113.84) node [anchor=north west][inner sep=0.75pt]   [align=left] {Ag. 2};
		% Text Node
		\draw (270,116.84) node [anchor=north west][inner sep=0.75pt]   [align=left] {Ag. 3};
		% Text Node
		\draw (419,115.84) node [anchor=north west][inner sep=0.75pt]   [align=left] {Ag. 5};
		% Text Node
		\draw (343,115.84) node [anchor=north west][inner sep=0.75pt]   [align=left] {Ag. 4};
		% Text Node
		\draw (296,53.84) node [anchor=north west][inner sep=0.75pt]   [align=left] {$\displaystyle \mathcal{T} = \text{Ag. 1}$};
		% Text Node
		\draw (298,194.84) node [anchor=north west][inner sep=0.75pt]   [align=left] {$\displaystyle \mathcal{R}= \text{Ag. 6}$};
	\end{tikzpicture}
	\caption{\label{scheme1}Schematic representation of the network composed of three layers. The top and bottom layers consist of only one agent (Agent 1, $\cal{T}$,  the transmitter, and Agent 6, $\cal{R}$, the receiver, respectively). The middle layer is composed of four agents interacting with the top and bottom layers. Links between the various agents are also shown. }
\end{figure}

In this Section, we suppose that information is conveyed from an initial transmitter to a final receiver through sub-agents. The scheme of the network is depicted in Figure~\ref{scheme1} and consists of three distinct layers, with the first layer being the transmitter, Agent 1 ($\cal{T}$), and the final layer being the receiver, Agent 6 ($\cal{R}$). The middle layer is composed of several agents that act as mediators between the transmitter and the receiver. We imagine that the flow of information follows a one-directional path, starting from the transmitter $\cal{T}$ and ending at the receiver $\cal{R}$, passing through the middle layer (this point, however, will be adequately relaxed adopting the ``Hermitian'' approach with the aid of the rules).

Given our focus on the duality of information as either \emph{good} or \emph{fake} (or pros and cons), we assume, as previously explained in Section \ref{sec::sec2},  that each agent in the network is connected to two separate fermionic modes, one of which is responsible for the transmission of good information and the other one for fake information. This means that the entire network involves 12 fermionic modes.
Moreover, we assume that the agents in the middle layer, namely Agents 2--5, have different behaviors: Agents 2 and 5 receive the news from $\cal T$ and only send the good and fake parts of the news, respectively. Conversely, Agent 3 and Agent 4 have the ability to modify the way in which news is transmitted to the lowest layer, changing  the nature of the information received from good to fake and vice versa, respectively. This is an important aspect of our model, as it adds an element of complexity to the transmission process. Overall, our analysis aims to shed light on the intricate ways in which information is conveyed through complex networks, with a particular focus on the role of individual agents in shaping the final outcome of the transmission process.

For such a model, we specialize the Hamiltonian operator $H$ in Equation (\ref{25}); namely, we consider
\begin{equation}
\label{ourHam}
\begin{aligned}
H &= \sum_{\alpha=1}^6\left(\omega_{f,\alpha}f^\dagger_\alpha f_\alpha+\omega_{g,\alpha}g^\dagger_\alpha g_\alpha\right)
+\sum_{\alpha\in\{1,3,4,6\}}\lambda_\alpha(f_\alpha g^\dagger_\alpha+g_\alpha f^\dagger_\alpha)\\
&+ \sum_{\beta=2}^5 \left(p^{(f)}_{1,\beta}(f_1f^\dagger_\beta+f_\beta f^\dagger_1)+
p^{(g)}_{1,\beta}(g_1g^\dagger_\beta+g_\beta g^\dagger_1)\right)\\
&+\sum_{\beta\in\{3,4,5\}} \left(p^{(f)}_{\beta,6}(f_\beta f^\dagger_6+f_6 f^\dagger_\beta)\right)
+\sum_{\beta\in\{2,3,4\}} \left(p^{(g)}_{\beta,6}(g_\beta g^\dagger_6+g_6 g^\dagger_\beta)\right),
\end{aligned}
\end{equation}
where all the parameters are real and positive.

The Hamiltonian Equation \eqref{ourHam} involves several parameters, and different choices for the values of the parameters, as well as the initial conditions, describe different situations and, not surprisingly, give different results.

Hereafter, we show the results of the numerical solutions by fixing the following set of~parameters:
\begin{equation}
\label{params}
\begin{aligned}
&\omega^{(f)}_1=1,\quad \omega^{(f)}_2=1.2,\quad  \omega^{(f)}_3=\omega^{(f)}_4=\omega^{(f)}_6=0.6,\quad  \omega^{(f)}_5=0.8,\\
&\omega^{(g)}_1=1,\quad \omega^{(g)}_2=\omega^{(g)}_3=\omega^{(g)}_4=0.8,\quad  \omega^{(g)}_5=1.2,\quad  \omega^{(g)}_6=0.6,\\
&\lambda_1=\lambda_3=\lambda_4=\lambda_6=0.2,\\
&p^{(f)}_{1,2}= p^{(f)}_{1,3}=p^{(f)}_{1,4}=p^{(f)}_{4,6}=0.5,\quad p^{(f)}_{1,5}=0.7,\quad p^{(f)}_{3,6}=0.3,\quad p^{(f)}_{5,6}=0.9,\\
&p^{(g)}_{1,2}=0.7,\quad p^{(g)}_{1,3}= p^{(g)}_{1,4}=p^{(g)}_{1,5}=p^{(g)}_{3,6}=0.5,\quad p^{(g)}_{2,6}=0.9,\quad p^{(g)}_{4,6}=0.3.
\end{aligned}
\end{equation}
Some comments on the rationale for this choice are in order:
\begin{itemize}
\item the inertia parameters associated with Agent 1 are equal (a form of neutrality of the transmitter); Agent 2 (Agent 5, respectively) has the higher value of the inertia parameter for fake news (for good news, respectively); for the remaining agents, the inertia parameters associated with fake news are smaller than those associated with good~news; 
\item the interaction parameters responsible for the conversion of good news into fake news and vice versa are all equal, except for Agents 2 and 5, where this interaction is~absent;
\item the coefficients related to the diffusion of good and fake news between the transmitter and the agents in the intermediate layer are all equal, except the coefficients involving Agents 2 and 5.
\end{itemize}

The main idea behind this particular choice of the values of the parameters is that, in general, fake news is more volatile than
good news; moreover, Agent 2 exchanges with the receiver only good news, whereas Agent 5 transmits to the receiver only fake news; finally, Agent 3 (Agent 4, respectively) conveys to the receiver more fake news than good news (more good news than fake news, respectively). 

Additionally, we choose the following initial values for the mean values:
\begin{equation}
\label{init}
F_1(0)=G_1(0)=\frac{1}{2}, \quad F_\alpha(0)=G_\alpha(0)=0, \quad \alpha=2,\ldots,6,
\end{equation}
which describe the fact that, at $t=0$, only agent 1 possesses some information, but its nature is not clear.

\begin{figure}[h]
{\includegraphics[width=0.49\textwidth]{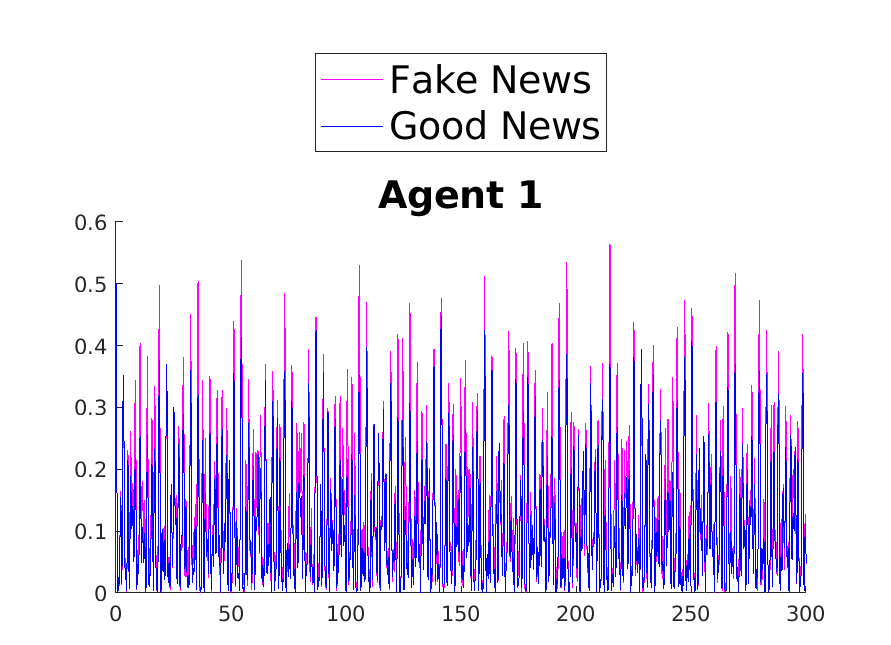}}
{\includegraphics[width=0.49\textwidth]{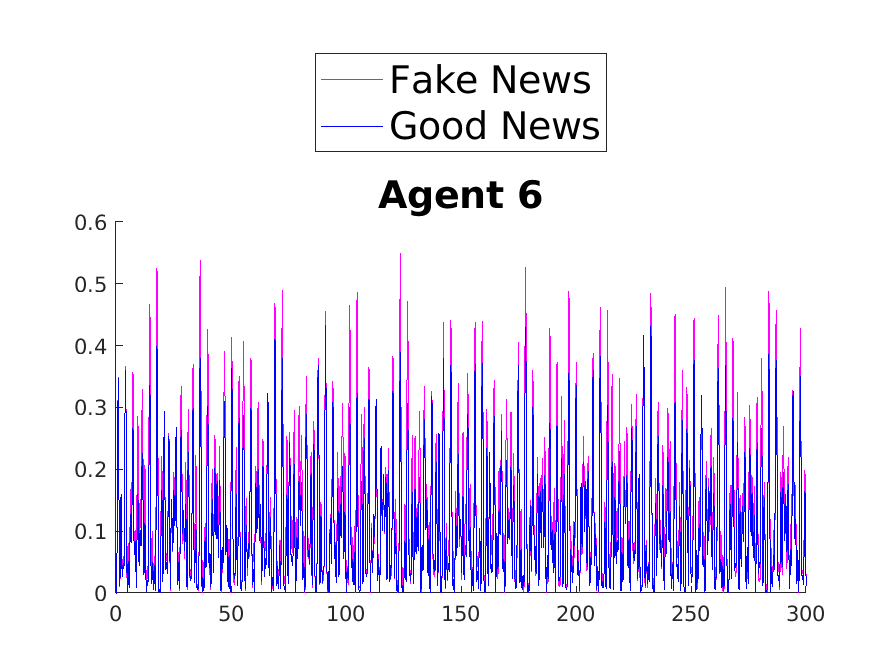}}\\
{\includegraphics[width=0.49\textwidth]{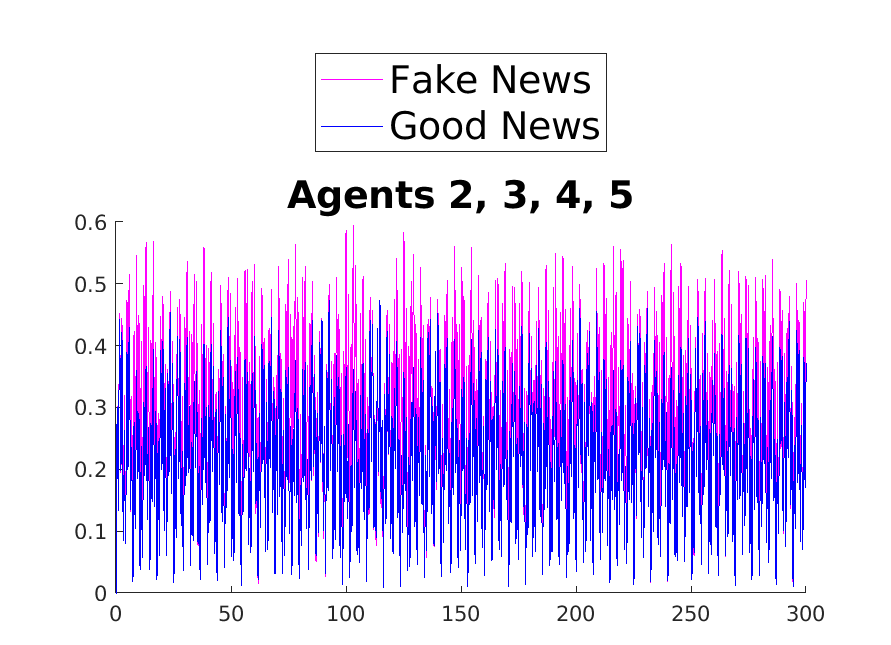}}
{\includegraphics[width=0.49\textwidth]{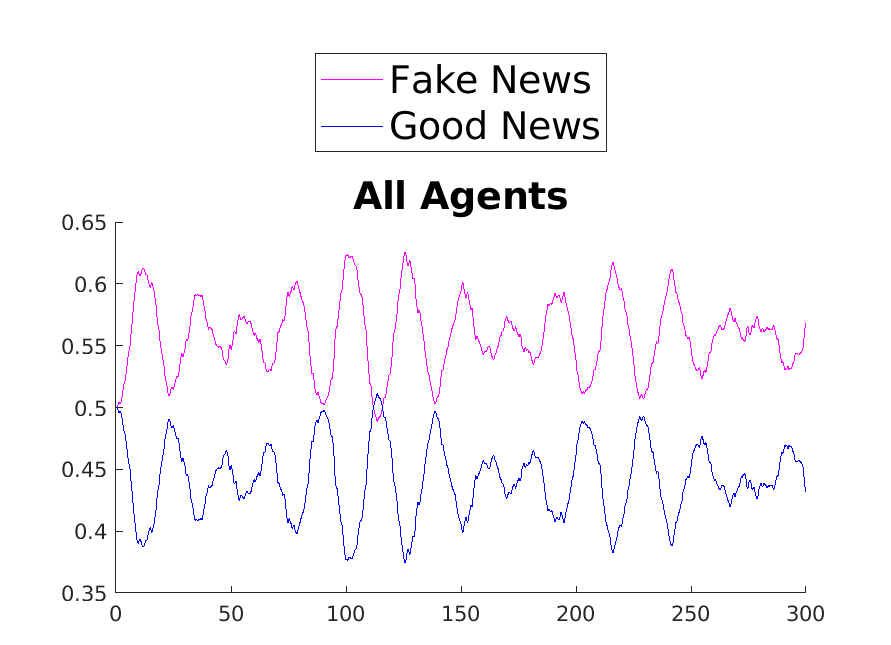}}
\caption{\label{fig:norule}Time evolution of the mean values of fake and good news for Agent 1 (top left), Agent 6 (top right), agents of the middle layer (bottom left), and all agents (bottom right).}
\end{figure}

The quadratic nature and the hermiticity of Equation \eqref{ourHam}, combined with the finite dimensionality of the system, imply that the solution of the differential equations derived in the Heisenberg view is, in general, quasiperiodic. This is evident in Figure~\ref{fig:norule}. 

In what follows, we switch to the $(H,\rho)$-induced dynamics (see \cite{HRO2} for details of the procedure). In particular, 
the  \emph{rule} $\rho$ that we introduce prescribes that the inertia parameters' $\omega^{(f)}$s and $\omega^{(g)}$s are periodically modified as a consequence of the evolution of the system; thus, the model adjusts itself during the time evolution, and this self-modification may be thought of as a surreptitious way to account for the influence of the external world on the attitudes of the various agents. 

Essentially, we start with the Hamiltonian Equation \eqref{ourHam}, where we fix the values of the parameters therein involved, and we compute, in a time interval of length $\tau>0$, the evolution of annihilation and creation operators, whereupon, choosing an initial condition for the mean values of the number operators, we obtain their time evolution (our observables). The variations in the values of the observables in the time interval $[0,\tau]$ determine, in a way that we describe below, a change in the inertia parameters. Then, a new Hamiltonian with the same functional structure but (possibly) different parameters arises, and we follow the continuous evolution of the system under the action of this new Hamiltonian for the next time interval of length $\tau$, i.e., for $t\in[\tau,2\tau[$,  and so on. In fact, taking a time interval $[0,T]$ where we study the evolution of the system, and splitting it into $n = T/\tau$ ($n$ is supposed, without loss of generality, to be an integer) subintervals of length $\tau$, we construct a sequence of Hamiltonians, differing from each other only in the values of some parameters; the global dynamics in the whole time interval $[0,T]$ is obtained 
by gluing the local evolutions in each subinterval. The result of this procedure is what we call $(H,\rho)$-induced dynamics. The choice of $\tau$ plays a role in the dynamics too. Indeed, if $\tau$ is very large, 
$\tau\simeq T$, it is as if no rule $\rho$ is truly acting on $\cal S$; vice versa, if $\tau\simeq0$, it is as if $\rho$ is acting continuously on $\cal S$ \cite{DSGO2020}, producing a form of Zeno effect.

{It is worth stressing that there	
	is no quantum feedback control in the approach of  $(H,\rho)$-induced dynamics, except that at a qualitative level. The main technical reason for the use of a rule $\rho$ relies on the need to have, in each time subinterval, time-independent linear Heisenberg equations, so as to limit the computational complexity for systems involving a large number of fermionic or bosonic modes. This approach allows us to modify some of the parameters entering the Hamiltonian according to the evolution of the system. In \cite{HRO2}, we showed that this approach can provide similar results to the ones derived when considering open quantum systems or in the case where the Hamiltonian is time-dependent (see also~\cite{DSGO2020}). The remarkable effect of the approach, besides keeping low the computational complexity, is that of producing time evolutions usually approaching some equilibrium state, even for finite-dimensional systems. Then, the  $(H,\rho)$-induced dynamics can be seen as an alternative to other existing methods usually adopted in the literature to describe systems that, after some transient, reach an equilibrium. }

To introduce our rules, let us define
\[
\begin{aligned}
&\delta^{(f)}_{\alpha}=F_\alpha(k\tau)-F_\alpha((k-1)\tau), \qquad &&\alpha=1,\ldots,6,\\
&\delta^{(g)}_{\alpha}=G_\alpha(k\tau)-G_\alpha((k-1)\tau), \qquad &&\alpha=1,\ldots,6,
\end{aligned}
\]
which are a measure of the variation in the fake and good  news perceived by Agent $A_\alpha$ in a time interval. These variations are used to propose six different rules, by updating at fixed instants $k\tau$ ($k=1,2,\ldots)$ the inertia parameters, mimicking in this way different variations in the attitudes of the agents of the network.
\begin{description}
\item[Rule 1:]%MDPI: Please confirm if the bold should be retained.
% Answer: Yes, the bold must be retained.

\[
\begin{aligned}
&\omega^{(f)}_\alpha \rightarrow \omega^{(f)}_\alpha(1+\kappa_\alpha\delta^{(f)}_{\alpha}),\\
&\omega^{(g)}_\alpha \rightarrow \omega^{(g)}_\alpha(1+\frac{\kappa_\alpha}{2}\delta^{(g)}_\alpha);
\end{aligned}
\]

\item[Rule 2:]
\[
\begin{aligned}
&\omega^{(f)}_\alpha \rightarrow \omega^{(f)}_\alpha(1+\frac{\kappa_\alpha}{2}\delta^{(f)}_\alpha),\\
&\omega^{(g)}_\alpha \rightarrow \omega^{(g)}_\alpha(1+\kappa_\alpha\delta^{(g)}_\alpha);
\end{aligned}
\]

\item[Rule 3:]
\[
\begin{aligned}
&\omega^{(f)}_\alpha \rightarrow \omega^{(f)}_\alpha(1+\kappa_\alpha\delta^{(f)}_\alpha),\\
&\omega^{(g)}_\alpha \rightarrow \omega^{(g)}_\alpha(1+\frac{\kappa_\alpha}{2}(\delta^{(g)}_\alpha)^2);
\end{aligned}
\]

\item[Rule 4:]
\[
\begin{aligned}
&\omega^{(f)}_\alpha \rightarrow \omega^{(f)}_\alpha(1+\kappa_\alpha(\delta^{(f)}_\alpha)^2),\\
&\omega^{(g)}_\alpha \rightarrow \omega^{(g)}_\alpha(1+\frac{\kappa_\alpha}{2}\delta^{(g)}_\alpha);
\end{aligned}
\]

\item[Rule 5:]
\[
\begin{aligned}
&\omega^{(f)}_\alpha \rightarrow \omega^{(f)}_\alpha(1+\frac{\kappa_\alpha}{2}\delta^{(f)}_\alpha),\\
&\omega^{(g)}_\alpha \rightarrow \omega^{(g)}_\alpha(1+\kappa_\alpha(\delta^{(g)}_\alpha)^2);
\end{aligned}
\]

\item[Rule 6:]
\[
\begin{aligned}
&\omega^{(f)}_\alpha \rightarrow \omega^{(f)}_\alpha(1+\frac{\kappa_\alpha}{2}(\delta^{(f)}_\alpha)^2),\\
&\omega^{(g)}_\alpha \rightarrow \omega^{(g)}_\alpha(1+\kappa_\alpha\delta^{(g)}_\alpha).
\end{aligned}
\]
\end{description}

We see that, in all the rules above, some positive coefficients $\kappa_\alpha$ are involved; they serve to weight differently the variations in the mean values in determining the modifications of the inertia parameters.

Some comments about the different rules are in order. Rules 1 and 2 are such that the various inertia parameters increase (decrease, respectively) if the corresponding mean values of fake and good news increase (decrease, respectively)  in the subinterval of length $\tau$, the only difference being the choice of the respective weights.
Due to the interpretation of the inertia parameters, this means that an increase in fake (or good) news tends to lower the  tendency to change in fake (or good) news. 

Using rule 3 (rule 4, respectively), the variation in the inertia parameters of fake news (good news, respectively) has the same trend as the variation in the corresponding mean values of fake news (good news, respectively), whereas the inertia parameters of good news (fake news, respectively) increase regardless of the sign of the variation in the mean values of good news  (fake news, respectively). This is because a square appears in the~formulas.

Finally, rules 5 and 6 are similar to rules 3 and 4, except for the inverted choice of the weights. These rules are interesting, in comparison with rules 3 and 4, to understand the role of the weights in the above formulas.

The numerical values of the weights that we use in the numerical solutions are
\begin{equation}
\label{weights}
\kappa_1=1, \quad \kappa_2=\kappa_3=\kappa_4=\kappa_5=1.2,\quad \kappa_6=0.6.
\end{equation}
Therefore, the variation in the inertia parameters of the agents of the middle layer is the largest one, and the weight associated with the variation in the inertia parameters of the receiver is the smallest one. 

\subsection*{Numerical Simulations}
The evolution of the mean values of fake and good news in the simplified network strongly depends, besides the choice of the rule, on the initial values of the parameters involved in the Hamiltonian. The effect determined by the rules is emphasized using, in all the numerical simulations, the set of initial parameters given in Equation \eqref{params}.

\begin{figure}[H]
\centering %%\fulllength If there is a figure in wide page, please release command \centering
{\includegraphics[width=0.49\textwidth]{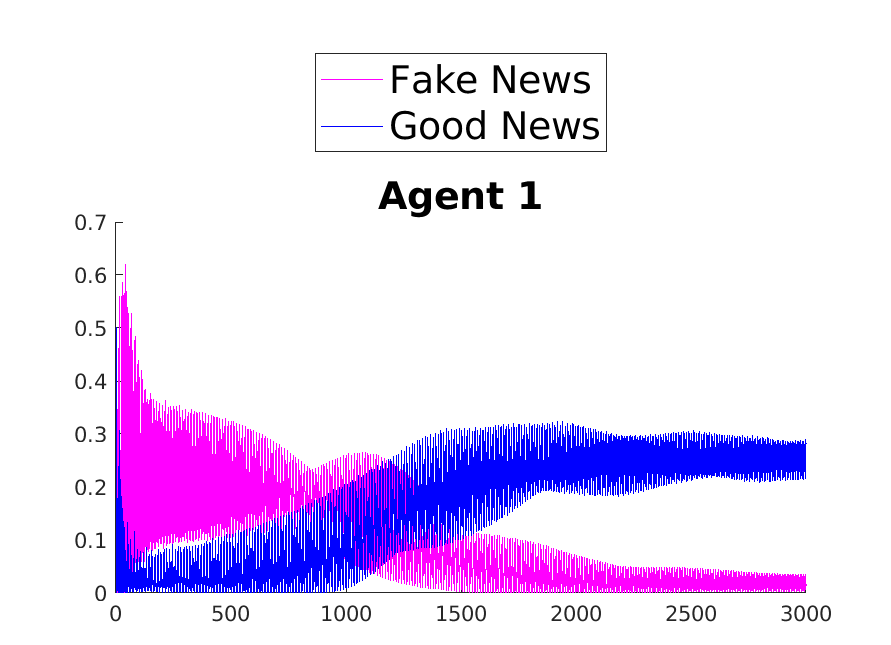}}
{\includegraphics[width=0.49\textwidth]{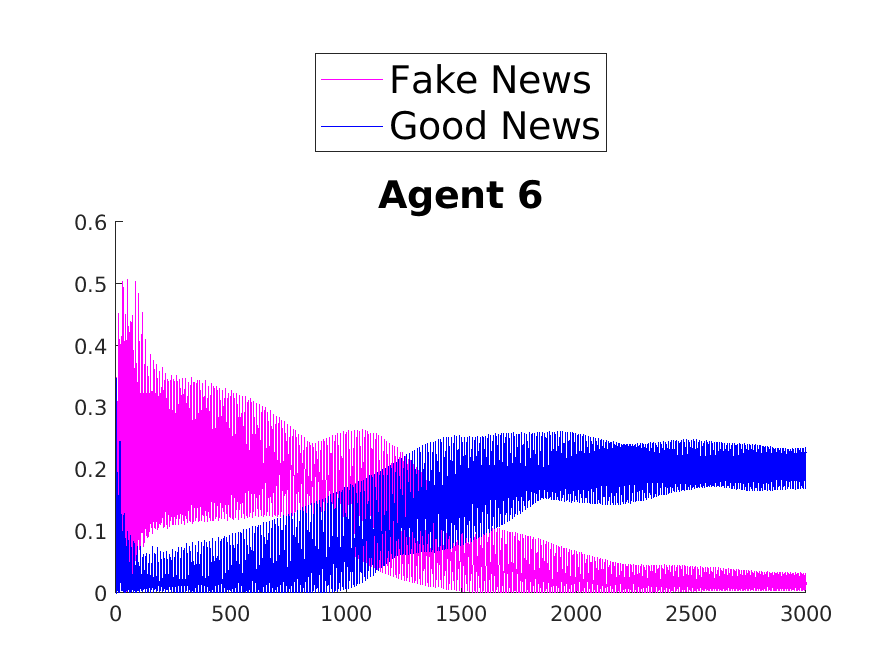}}\\
{\includegraphics[width=0.49\textwidth]{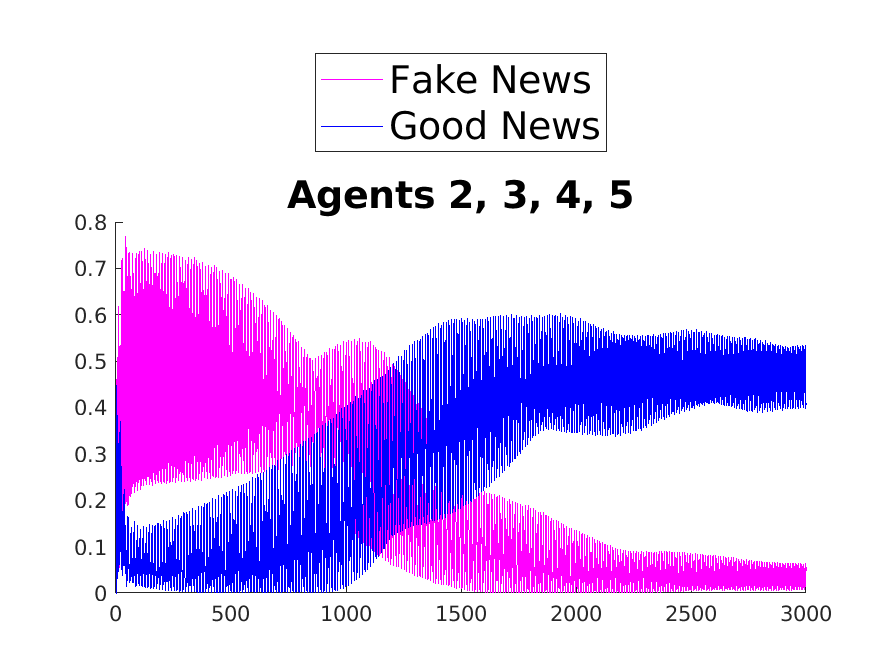}}
{\includegraphics[width=0.49\textwidth]{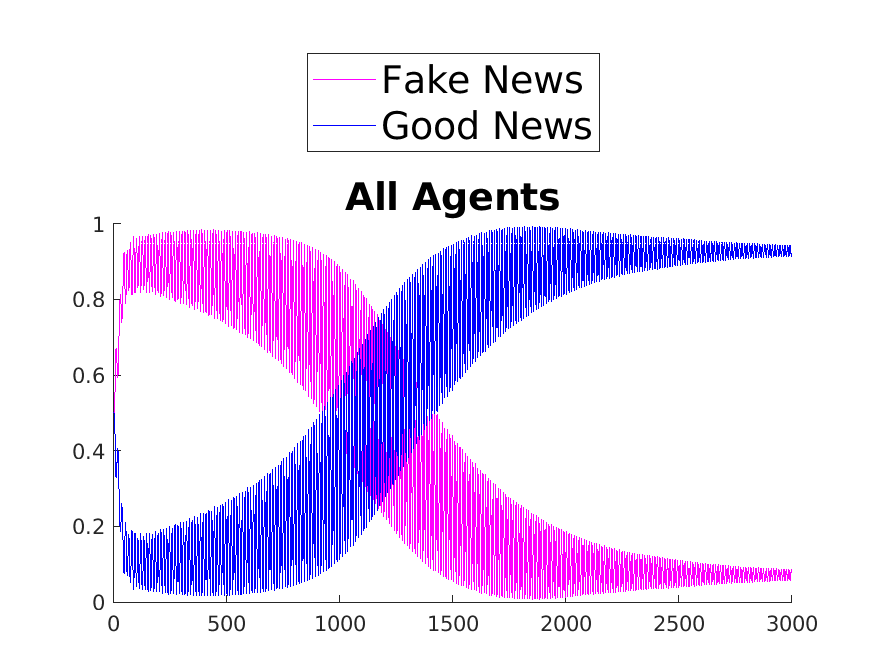}}
\caption{\label{fig:rule1}Time evolution of the mean values of fake and good news for Agent 1 (top left), Agent 6 (top right), agents of the middle layer (bottom left), and all agents (bottom right); rule 1 with $\tau=1$.}
\end{figure}

\begin{figure}[H]
\centering %%\fulllength If there is a figure in wide page, please release command \centering
{\includegraphics[width=0.49\textwidth]{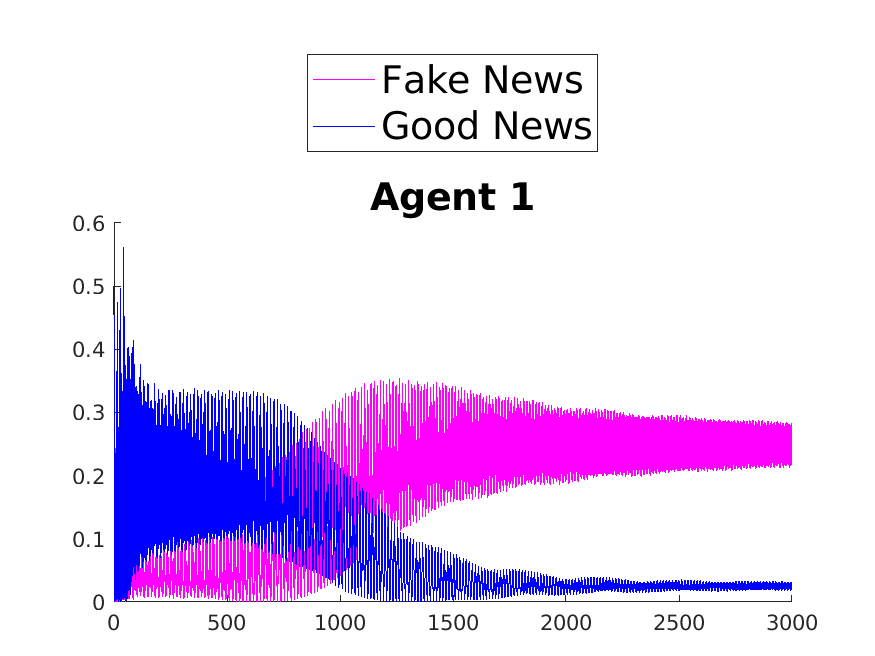}}
{\includegraphics[width=0.49\textwidth]{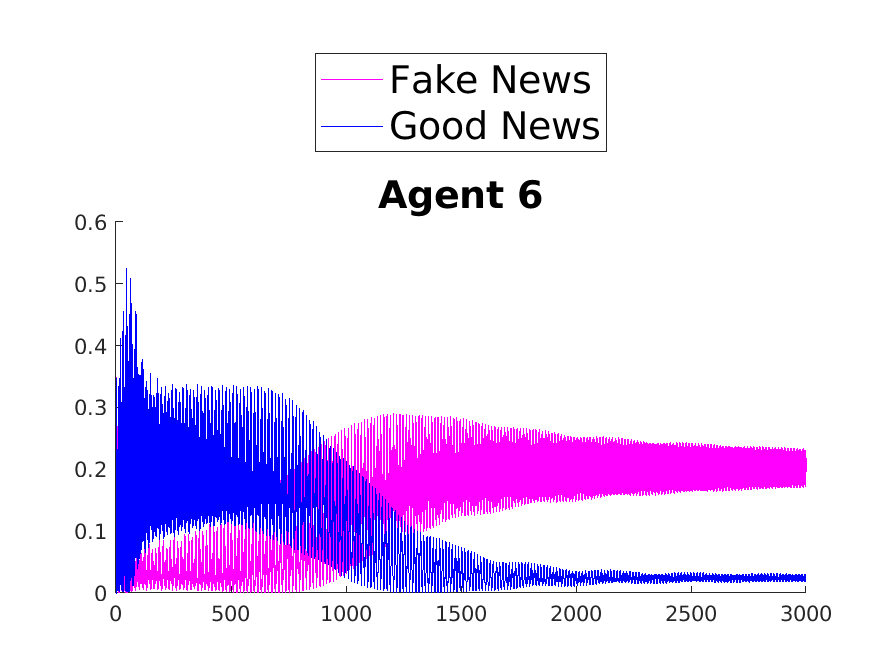}}\\
{\includegraphics[width=0.49\textwidth]{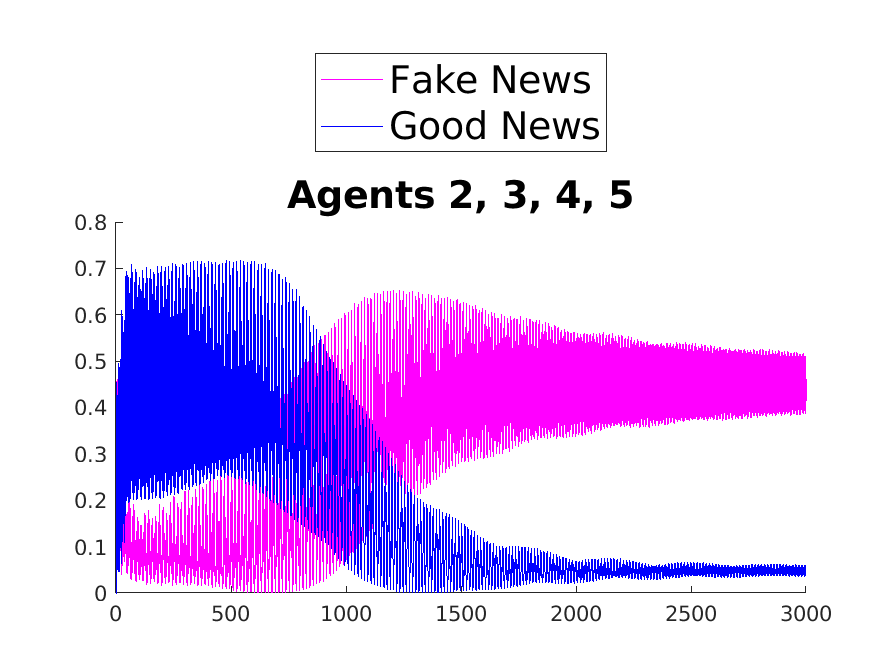}}
{\includegraphics[width=0.49\textwidth]{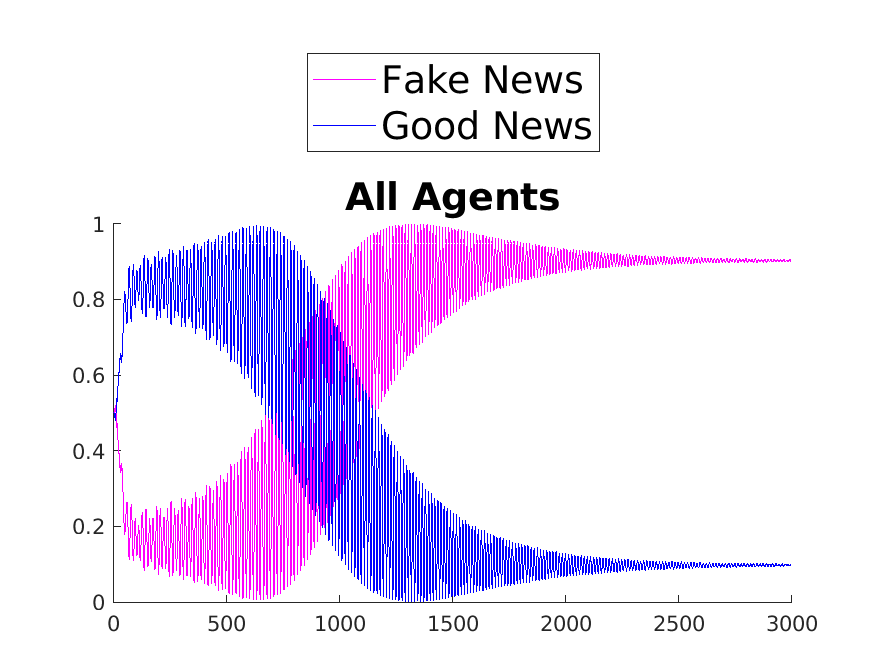}}
\caption{\label{fig:rule2}Time evolution of the mean values of fake and good news for Agent 1 (top left), Agent 6 (top right), agents of the middle layer (bottom left), and all agents (bottom right); rule 2 with $\tau=1$.}
\end{figure}

\begin{figure}[H]
\centering %%\fulllength If there is a figure in wide page, please release command \centering
{\includegraphics[width=0.49\textwidth]{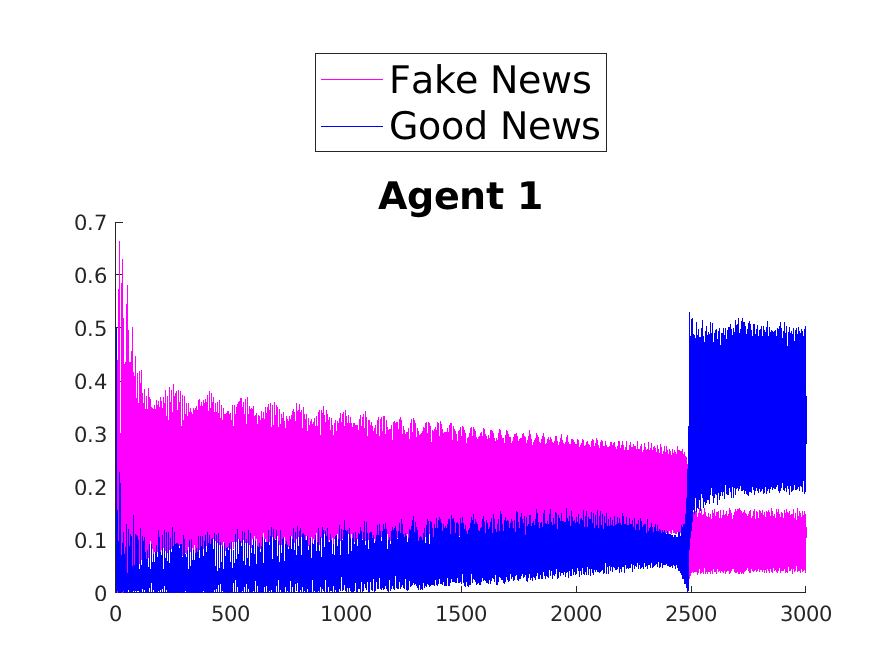}}
{\includegraphics[width=0.49\textwidth]{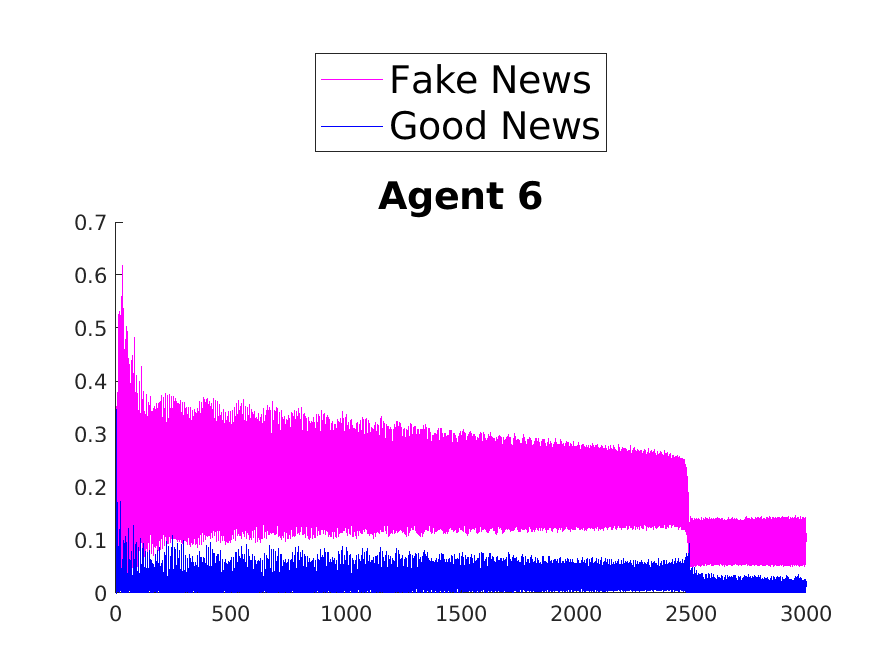}}\\
{\includegraphics[width=0.49\textwidth]{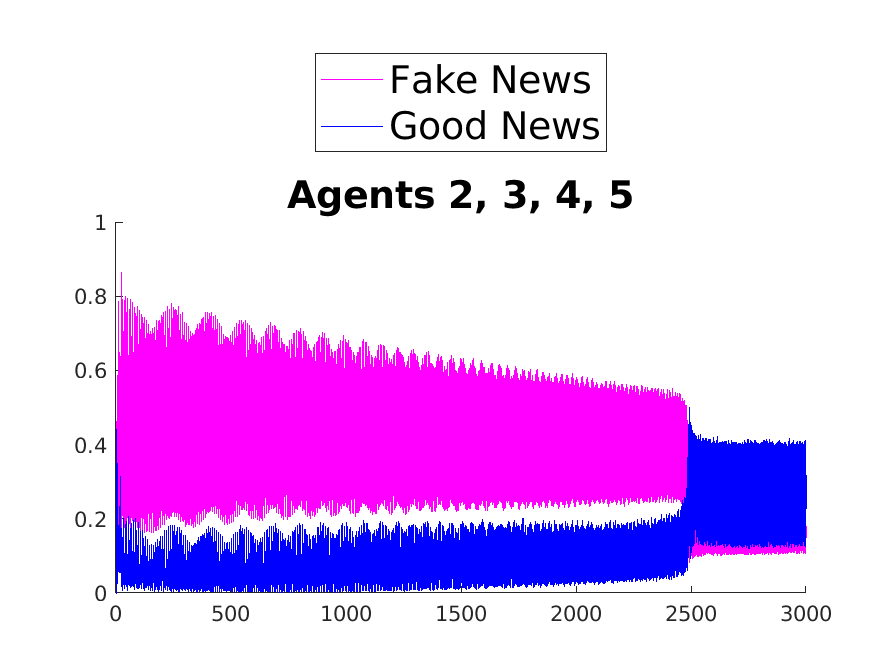}}
{\includegraphics[width=0.49\textwidth]{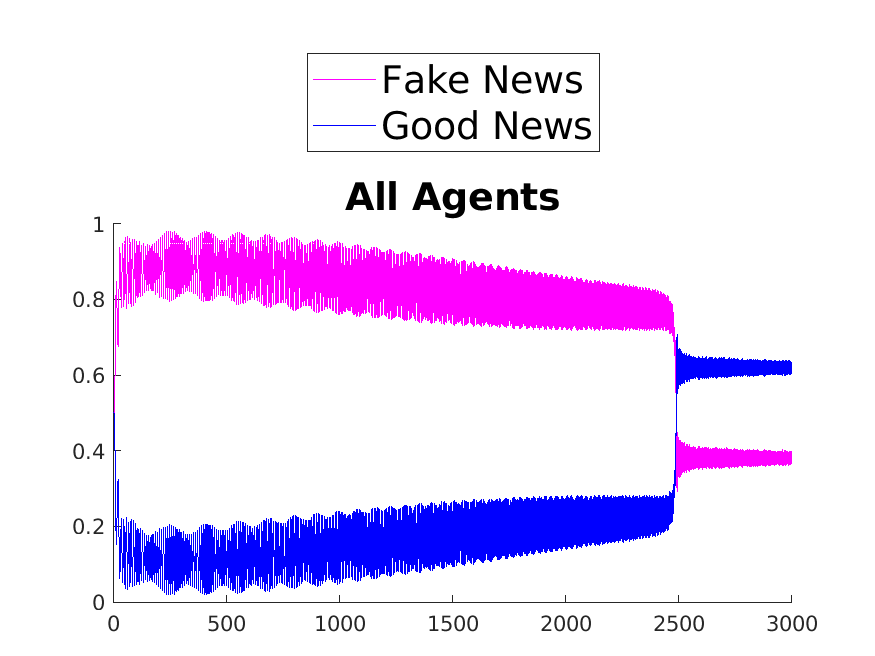}}
\caption{\label{fig:rule3}Time evolution of the mean values of fake and good news for Agent 1 (top left), Agent 6 (top right), agents of the middle layer (bottom left), and all agents (bottom right); rule 3 with $\tau=1$.}
\end{figure}

\begin{figure}[H]
\centering %%\fulllength If there is a figure in wide page, please release command \centering
{\includegraphics[width=0.49\textwidth]{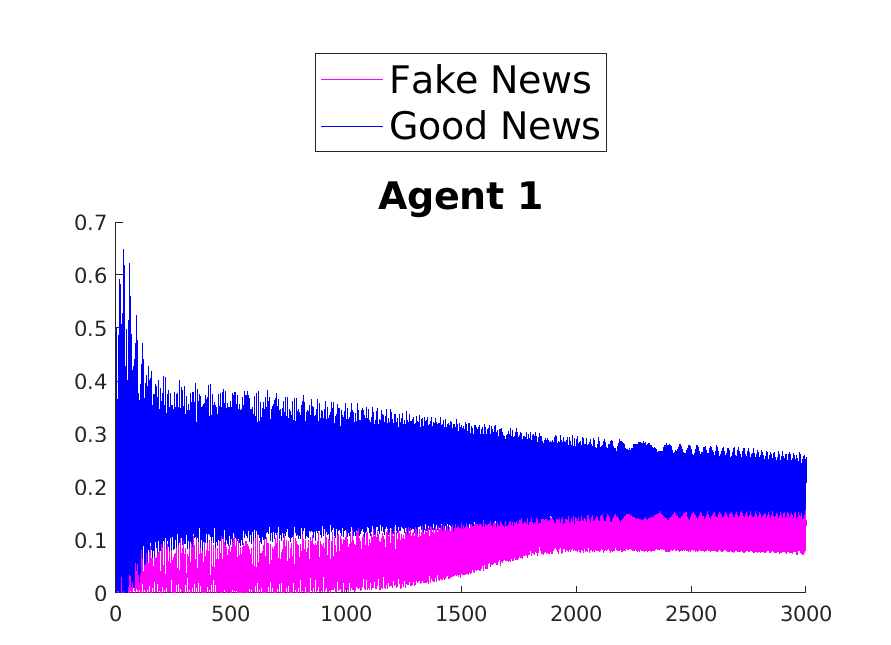}}
{\includegraphics[width=0.49\textwidth]{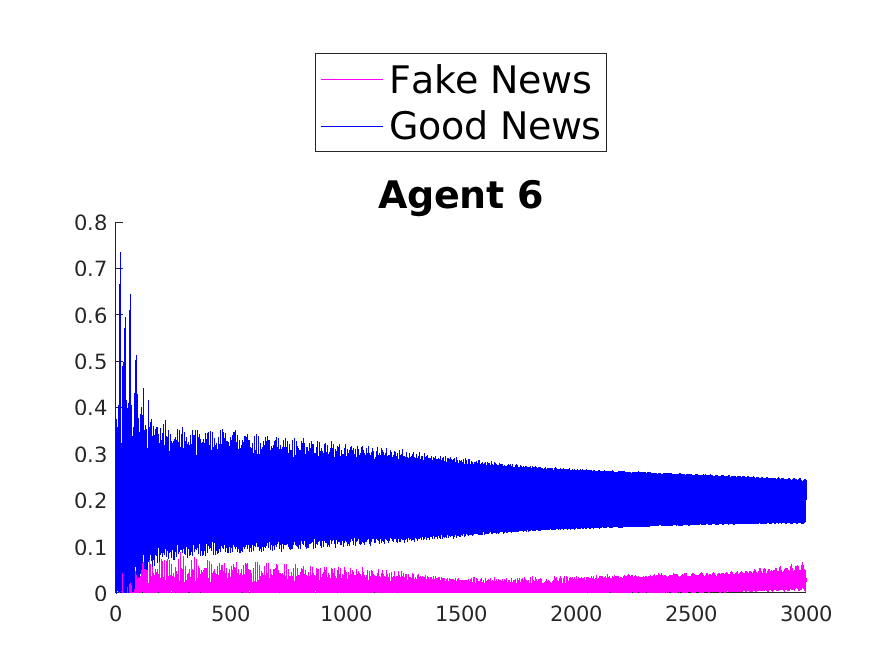}}\\
{\includegraphics[width=0.49\textwidth]{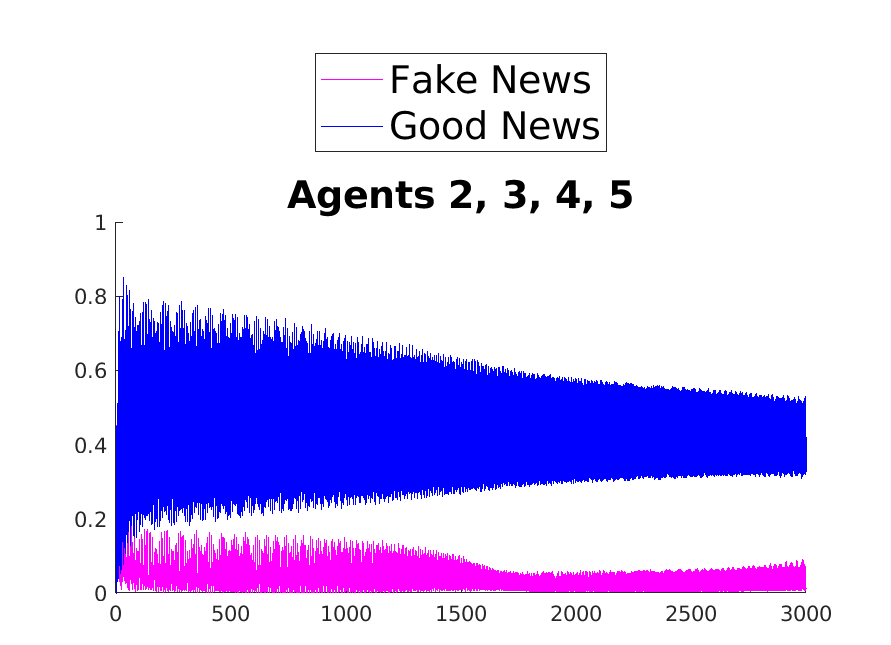}}
{\includegraphics[width=0.49\textwidth]{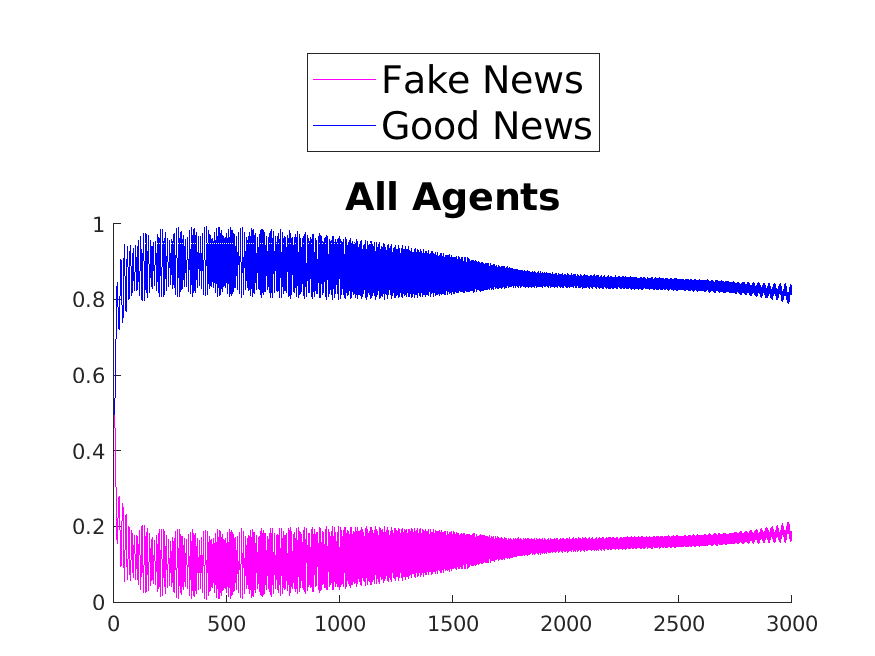}}
\caption{\label{fig:rule4}Time evolution of the mean values of fake and good news for Agent 1 (top left), Agent 6 (top right), agents of the middle layer (bottom left), and all agents (bottom right); rule 4 with $\tau=1$.}
\end{figure}

\begin{figure}[H]
\centering %%\fulllength If there is a figure in wide page, please release command \centering
{\includegraphics[width=0.49\textwidth]{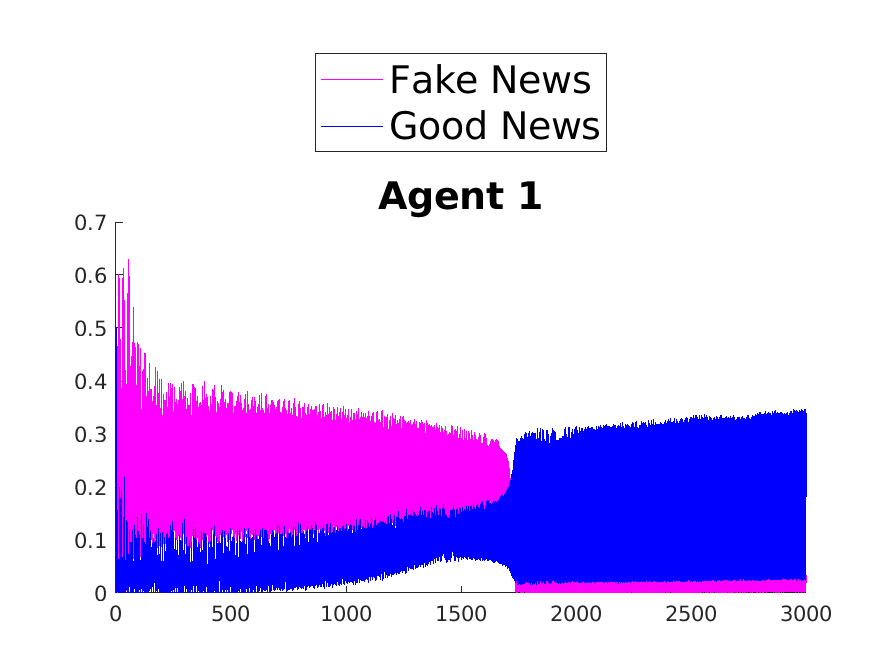}}
{\includegraphics[width=0.49\textwidth]{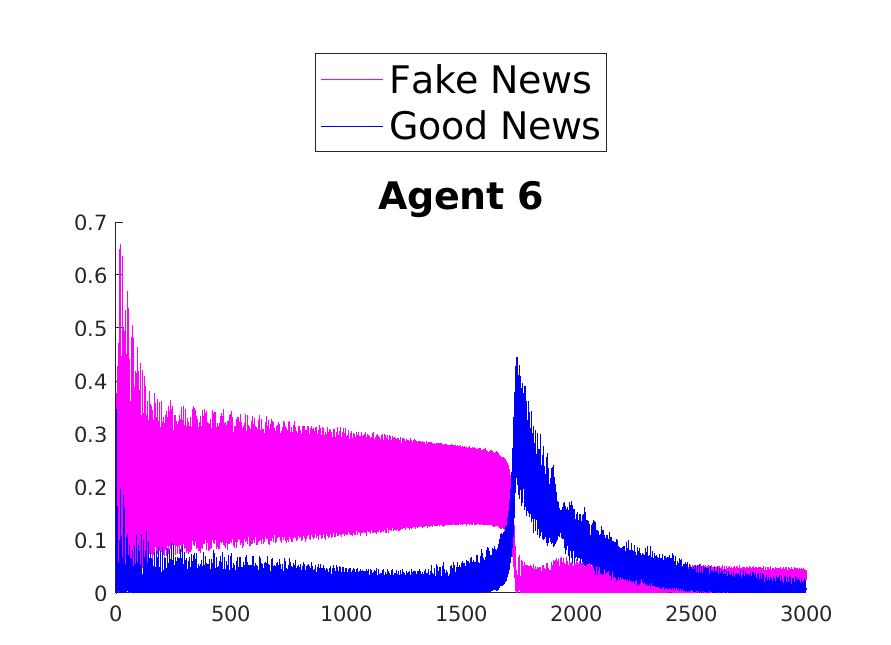}}\\
{\includegraphics[width=0.49\textwidth]{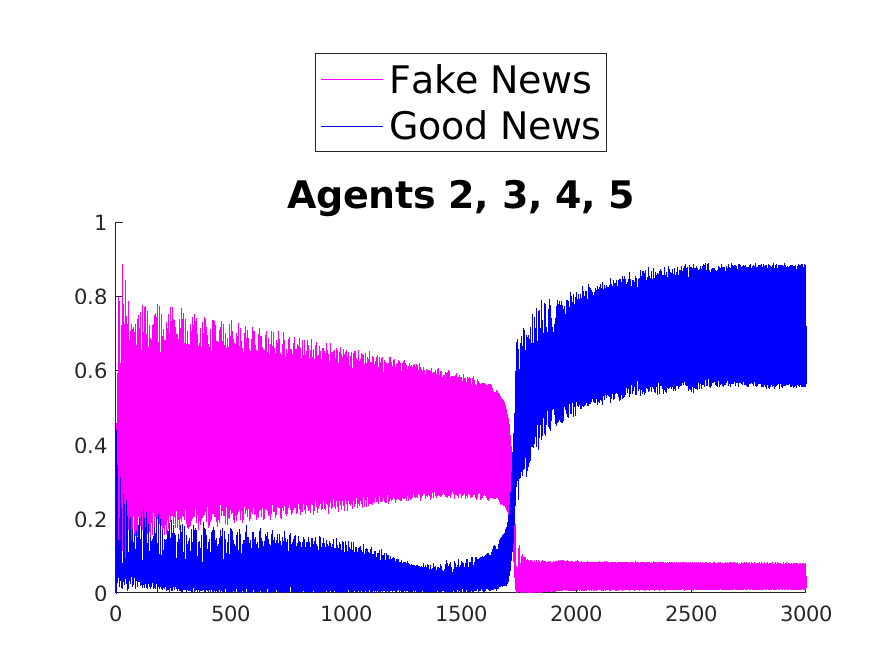}}
{\includegraphics[width=0.49\textwidth]{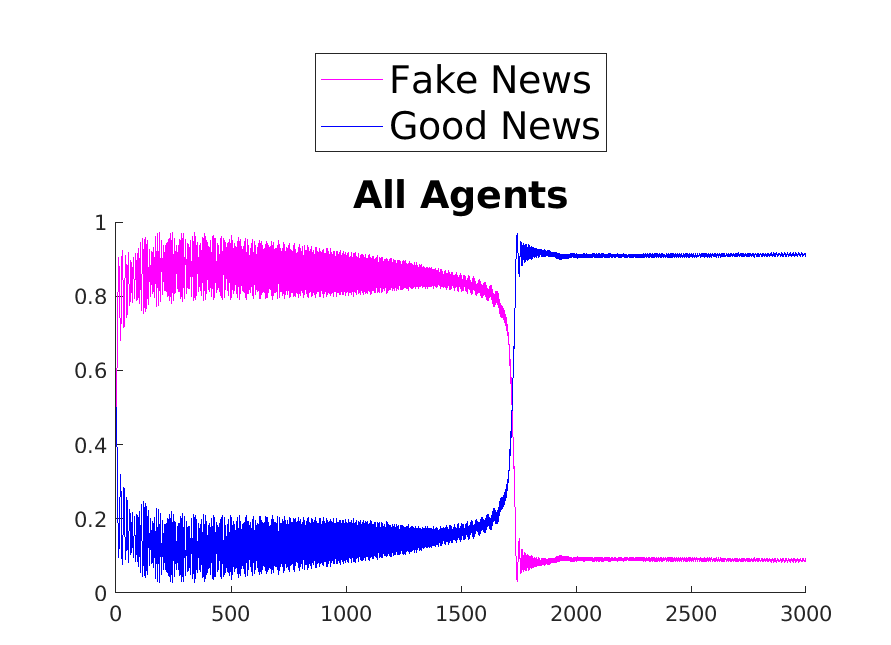}}
\caption{\label{fig:rule5}Time evolution of the mean values of fake and good news for Agent 1 (top left), Agent 6 (top right), agents of the middle layer (bottom left), and all agents (bottom right); rule 5 with $\tau=1$.}
\end{figure}

\begin{figure}[H]
\centering %%\fulllength If there is a figure in wide page, please release command \centering
{\includegraphics[width=0.49\textwidth]{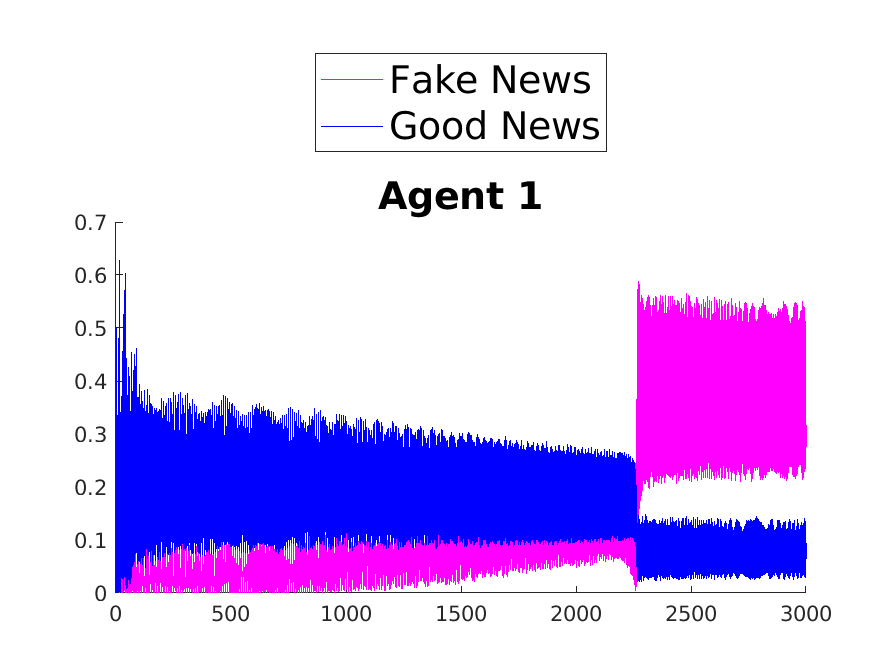}}
{\includegraphics[width=0.49\textwidth]{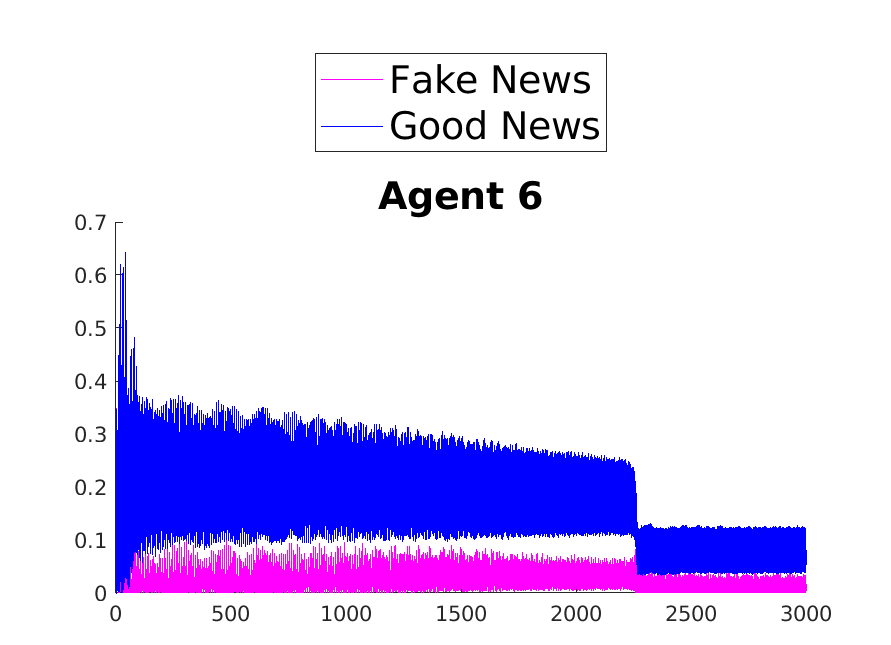}}\\
{\includegraphics[width=0.49\textwidth]{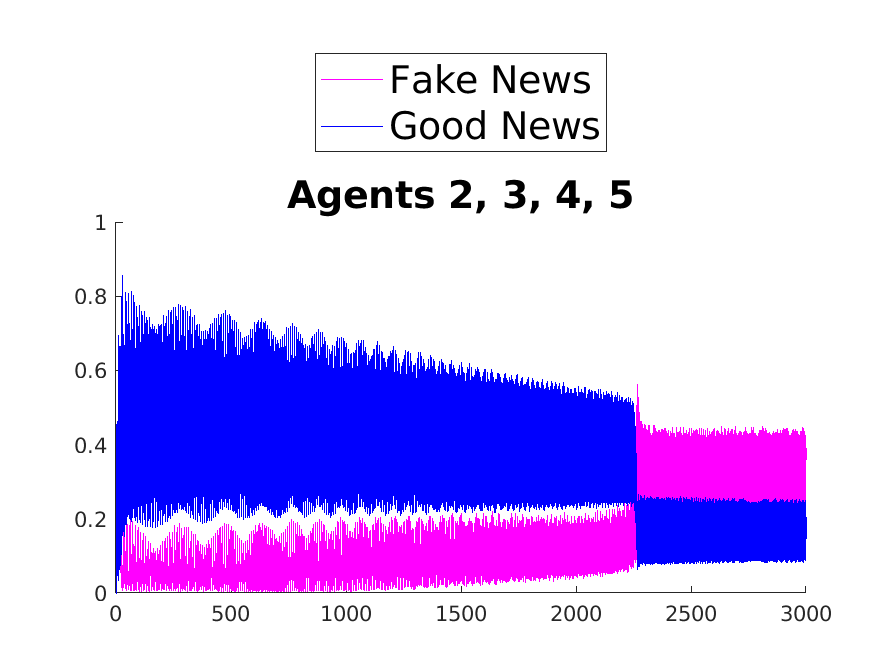}}
{\includegraphics[width=0.49\textwidth]{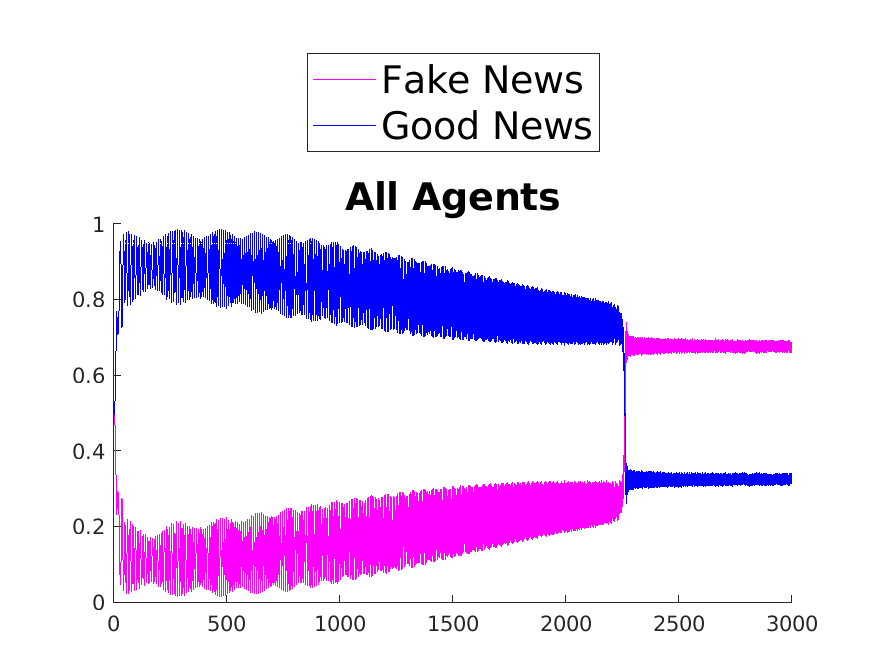}}
\caption{\label{fig:rule6}Time evolution of the mean values of fake and good news for Agent 1 (top left), Agent 6 (top right), agents of the middle layer (bottom left), and all agents (bottom right); rule 6 with $\tau=1$.}
\end{figure}

Figures~\ref{fig:rule1}--\ref{fig:rule6} show the time evolution of the news in the network according to the different rules.

What can be observed in all the figures is that different rules determine different outcomes, even if the initial condition and the initial values of the parameters are unchanged. The receiver can obtain a different combination of fake and good news, and this depends on the mechanisms changing the inertia parameters of the various agents. Moreover, the rules have the effect of introducing a form of irreversibility in the dynamics, despite the hermiticity of the Hamiltonian.

Among other effects, we observe that, adopting rule 1, 2, or 5, there results a change in what Agent 6 perceives after some time; on the contrary, using rules 3, 4, and 6, despite the presence of many oscillations, the overall {\em {impression}} of Agent 6 remains unchanged. On the other hand, what happens for the agents in the intermediate layer shows large variability, possibly due to the presence of agents with different attitudes in the layer. Quite interestingly, even if the news passes from Agent 1 to Agent 6, all the figures clearly show a {\em {feedback effect}} on the transmitter, which can change its perception of ${\cal N}$ with time. This is due to the oscillatory behavior of the dynamics in each intermediate time interval $[(k-1)\tau, k\tau]$, $k=1,2,\ldots$. This means that also the transmitter $\cal{T}$ can modify its  perception as a consequence of its interaction with the other agents. 

\section{Dynamics of the System Using the GKSL Equations}
\label{section::GKSL}
In this Section, we propose a different approach, based on the GKSL equation, to solve the dynamics within the network schematized in Figure~\ref{scheme1}. Recently, the GKSL has been used to model several types of macroscopic systems, particularly those requiring the preservation of irreversibility in certain mechanisms, as done in recent applications of the Lindblad approach beyond the quantum domain (see \cite{khren1, Asan2013, Nava2022, Nava2023, BaGa23} and the interesting discussions in \cite{Krea23}).

As highlighted previously, the network dynamics is based on information flowing unidirectionally among the agents. In fact, the dynamics can be visualized as a series of continuous transmissions of small packets of information from one agent to another. Therefore, we expect this mechanism to be accurately described only by the operator of the form $a_\alpha^\dag a_\beta$ (and not considering their adjoint), where the $\beta$-th agent is transmitting to the $\alpha$-th agent, and $a$ can be any of the $f$ or $g$ operators. Similarly, the process of changing the nature of the information for a generic $\alpha$-th agent can be described using operators of the form $g_\alpha^\dag f_\alpha$ or $f_\alpha^\dag g_\alpha$.

This suggests that the GKSL equation, which is well suited to solving dynamics where non-reversible effects are predominant, could be an appropriate approach to describing the dynamics within the network. The GKSL equation describes the evolution of the density operator $\rho(t)=|\Psi(t)\rangle\langle\Psi(t)|$ (written in the usual bra--ket notation), where $\Psi(t)$ represents the state of the system at time $t$, and reads as follows:
\begin{equation}
\frac{d}{d t}\rho(t)=-i [H, \rho(t)]+\sum_{\alpha=1}^{N^\star}\left(L_{\alpha}\, \rho(t) L_{\alpha}^{\dagger}-\frac{1}{2}\left\{L_{\alpha}^{\dagger} L_{\alpha}\, \rho(t)\right\}\right),
\label{eq:lineq}
\end{equation}
where $H$ is the Hamiltonian of the system, which typically contains only reversible effects, such as the ones in Equations (\ref{25}) and (\ref{ourHam}), for instance. The various $L_\alpha$s represent the Lindblad operators responsible for the irreversible mechanisms, and $N^\star$ is their number. When writing the GKSL equation, we assume that the system dynamics is Markovian, and the Lindblad operators are independent of the state of the system.

Compared to the previous approach, which was mostly based on the evolution of operators through the Heisenberg equation, the primary difference now is that the density operator connected to the state $\Psi$ is evolving. Using this approach, one can easily derive the mean values of the number operators, as shown in Equations \eqref{29} and \eqref{210}, using the standard procedure based on the computation of the traces:
\begin{equation}
G_\alpha(t)=\text{tr}(\rho(t)\,\widehat{G}_\alpha),\qquad F_\alpha(t)=\text{tr}(\rho(t)\,\widehat{F}_\alpha),\qquad \alpha=1,\ldots,N.
\end{equation}

To be more specific, we assume that $H$ is nothing but the free  and self-adjoint Hamiltonian $H_0$ in Equation \eqref{25},
\begin{equation}
H=H_0=\sum_{\alpha=1}^6 \omega^{(f)}_\alpha\widehat{F}_\alpha+\sum_{\alpha=1}^6\omega^{(g)}_\alpha\widehat{G}_\alpha,\label{l32}
\end{equation}
whereas the various Lindblad operators responsible for the transmission between the different layers are
\begin{eqnarray}
L_{\alpha-1}=L_{\alpha,1}^{(g)}&=&p^{(g)}_{\alpha,1}\,g_\alpha^\dag g_1,\quad\text{for } \alpha=2,\ldots, 5,\label{l33}\\
L_{\alpha+3}=L_{\alpha,1}^{(f)}&=&p^{(f)}_{\alpha,1}\,f_\alpha^\dag f_1,\quad\text{for } \alpha=2,\ldots, 5,\\
L_{\alpha+7}=L_{6,\alpha}^{(g)}&=&p^{(g)}_{6,\alpha}\,g_6^\dag g_\alpha,\quad\text{for } \alpha=2,\ldots, 5,\\ 
L_{\alpha+11}=L_{6,\alpha}^{(f)}&=&p^{(f)}_{6,\alpha}\,f_6^\dag f_\alpha,\quad\text{for } \alpha=2,\ldots, 5,\label{l36}
\end{eqnarray}
and those responsible for changing the nature of the information in the agents $3$ and $4$ are
\begin{equation}
L_{17}=p^{(g)}_{3,3}\,g_3^\dag f_3,\qquad L_{18}=p^{(f)}_{4,4}\,f_4^\dag g_4.\label{l37} 
\end{equation}
Understanding how the various Lindblad operators affect the system is quite straightforward. It is well known that any initial pure state $\Psi$ evolves into a mixture of states due to the occurrence of \textit{{evolutionary jumps}} \cite{Manz}. Similar computations to those performed in \cite{BaGa23} can be useful to explain this mixture and are based on a standard perturbative approach in Equation \eqref{eq:lineq}. In doing so, we can neglect the action of the Hamiltonian $H$, focusing only on the Lindblad operators. To the leading order in $dt$, the evolved density operator $\rho(dt)$ of a pure state in a small time step $dt$ is given by
\begin{equation}
\rho(dt)\approx 
\mathcal{A}\, \rho\,\mathcal{A}^\dagger+\sum_{\alpha=1}^{N^\star}\mathcal{B}_\alpha\, \rho\,\mathcal{B}_\alpha^\dagger, 
\label{linjumps}
\end{equation}
where 
\begin{equation}
\mathcal{A}=\1-\frac{dt}{2}\sum_{\alpha=1}^{N^\star}\tilde L_{\alpha}^{\dagger}\tilde L_{\alpha},\qquad
\mathcal{B}_\alpha=\sqrt{dt}\,\tilde L_\alpha,\quad \alpha=1,\ldots,N^\star;
\end{equation}
$N^\star=18$ is the total number of Lindblad operators that are defined in
Equations \eqref{l33}--\eqref{l37}.
This means that  the evolved state is a mixture of the pure states defined, with suitable normalizations, by
$\mathcal{A}\Psi$ and by the various $\mathcal{B}_\alpha\Psi$.
In particular, there is a probability 
\[
p_{\mathcal{A}}={\|\mathcal{A}\, \Psi\|}^2\simeq\left(1-dt\sum_{\alpha=1}^{N^\star}\|\tilde L_\alpha\Psi\|^2\right)
\] 
that the state $\Psi$ evolves following the \textit{{continuous drift-type evolution}} to 
\begin{equation}
\frac{1}{\|\mathcal{A}\, \Psi\|}\mathcal{A}\, \Psi,
\end{equation}
and there is a probability $p_{\mathcal{B}_\alpha}={\|\mathcal{B}_\alpha\, \Psi\|}^2=dt\|\tilde L_\alpha\Psi\|^2$ 
that evolves following the various \textit{{evolutionary jumps}} in 
\begin{equation}
\frac{1}{\|\mathcal{ B}_\alpha\Psi\|}\mathcal{B}_\alpha\Psi,\qquad \alpha=1,\ldots N^\star.
\end{equation}

In the latter process, the evolved state does not tend to the original one as $dt\rightarrow0$, as in the process ruled by the \textit{{continuous drift-type evolution}}, and this causes the well-known mixture of states. We remark that all the Lindblad operators in Equations \eqref{l33}--\eqref{l36}  are of the type  $a^\dag b$, where $a$ and $b$ can be  $g_{j}$ or $f_{k}$, so that the evolutionary jumps work by pushing the pure state
$\Psi=\sum\alpha_{\bf{n,m}}\Psi_{\mathbf{n,m}} $ with $\bf{ n}$ or $ \bf{m}$ (depending on whether we act with $g$ or $f$, respectively) of the form $(\ldots,\underbrace{0}_{j-\textrm{th}},\ldots,\underbrace{1}_{k-\text{th}},\ldots)$  into the state $\sum\alpha_{\bf{n,m}}\Psi_{\mathbf{n^\star,m^\star}} $,
where  $\bf{n^\star}$ or $\bf{m^\star}$ has the form $(\ldots,\underbrace{1}_{j-\textrm{th}},\ldots,\underbrace{0}_{k-\text{th}},\ldots)$, while
 the components  different from $j,k$ are the same as those of $\bf{n},\bf{m}$. For instance, the action of $\tilde L_1=L_{2,1}^{(g)}=p^{(g)}_{2,1}\,g_2^\dag g_1$ causes a jump from the state with $\bf{n}=(1,0,\ldots)$ to a state with $\bf{n}^\star=(0,1,\ldots)$. Similarly, if we act with one of the  operators in Equation \eqref{l37}, let us say $L_{17}$,  we would have that the pure state
 $\Psi=\sum\alpha_{\bf{n,m}}\Psi_{\mathbf{n,m}} $ with $\bf{ n}$ and $ \bf{m}$ of the form $(\ldots,\underbrace{0}_{3-\textrm{th}},\ldots)$ or $(\ldots,\underbrace{1}_{3-\textrm{th}},\ldots)$ drifts into the state $\sum\alpha_{\bf{n,m}}\Psi_{\mathbf{n^\star,m^\star}} $,
 where  $\bf{n^\star}$ and $\bf{m^\star}$ have the form $(\ldots,\underbrace{1}_{3-\textrm{th}},\ldots)$ or $(\ldots,\underbrace{0}_{3-\textrm{th}},\ldots)$, respectively.
 
From this qualitative analysis, it is expected that, in our model, all the Lindblad operators operate in a chain driving the quantity of information contained in the first layer toward the last layer. Hence, as time goes on, the values of $G_1(t)$ and $F_1(t)$ in the first layer will decrease, while the other layers will increase their relative values. Ultimately, all the information will be contained in $G_6(t)$ and $F_6(t)$, depending on the strength of the layers responsible for switching the information (in particular, Agents 3 and 4).  In the following, we consider three different numerical experiments intended to show how it is possible to simulate different situations in which the news perceived by $\cal R$ can be dependent on the mechanisms ruling the transmission in the middle layer.

 \subsection{Experiment I}
\begin{figure}[H] 
 \centering
{\captionsetup{position=bottom,justification=centering}
 \subfloat[]{\includegraphics[width=0.49\textwidth]{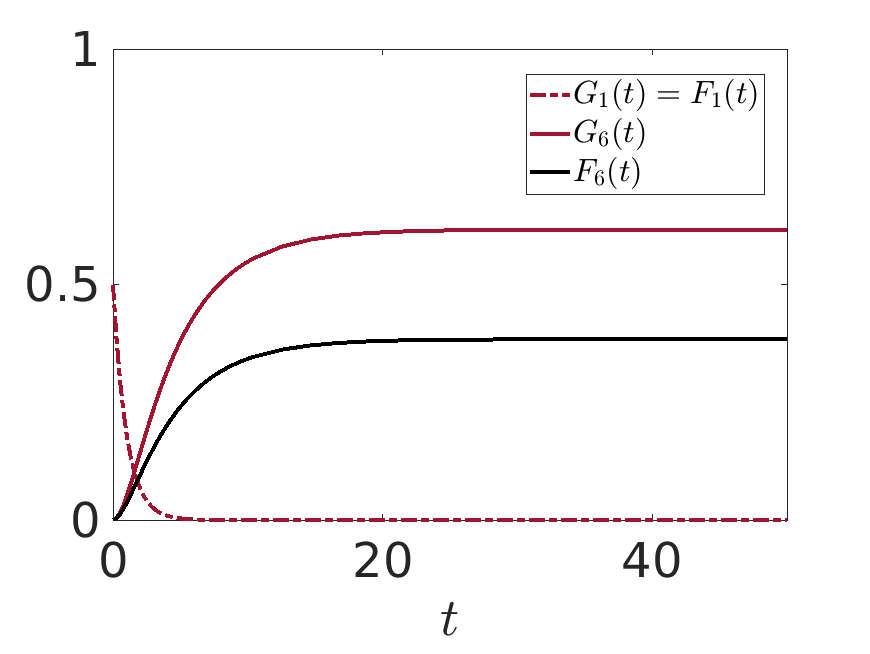}}
\subfloat[]{\includegraphics[width=0.49\textwidth]{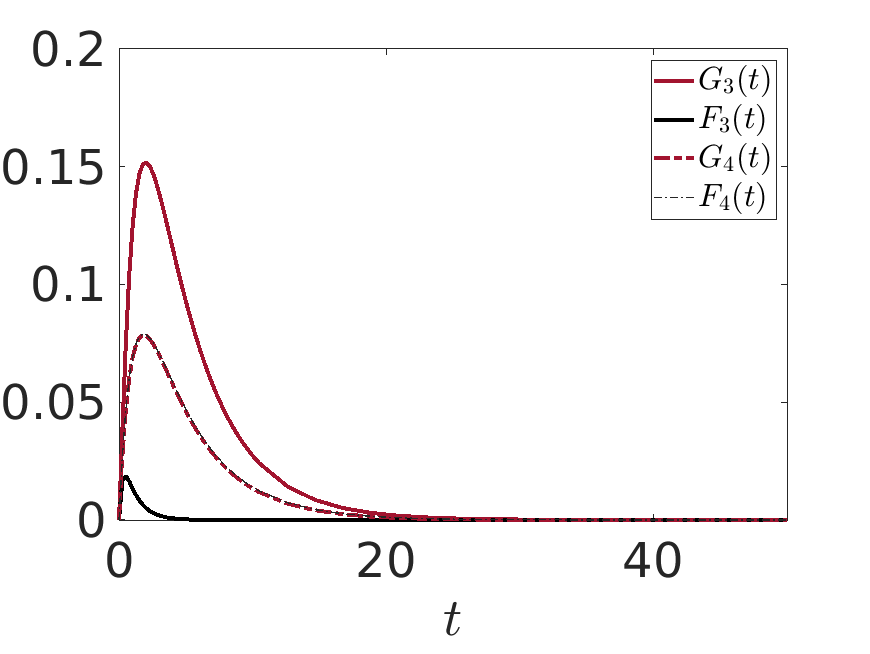}}\\
\subfloat[]{\includegraphics[width=0.49\textwidth]{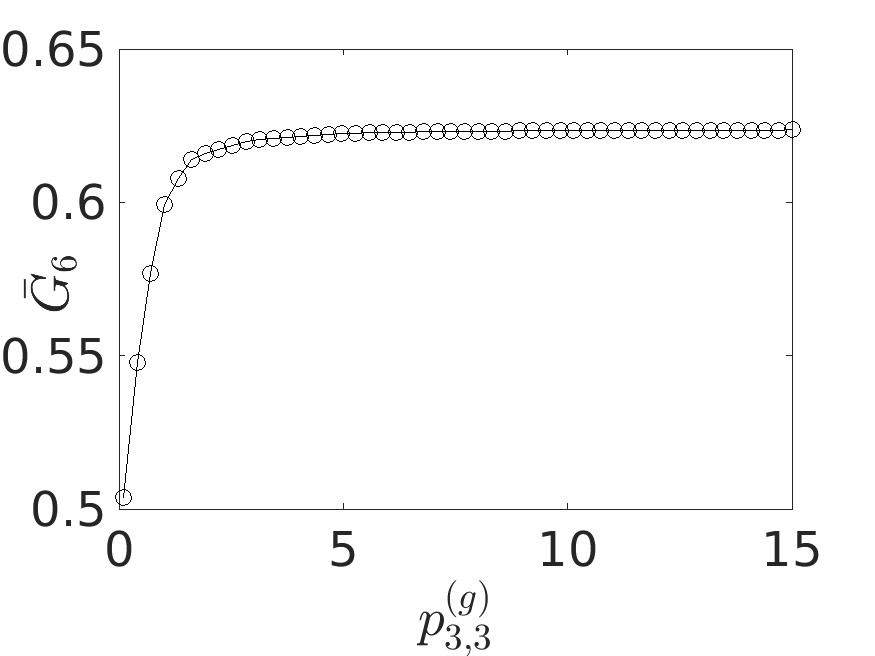}}
}
 \caption{Time evolutions of (\textbf{a}) the mean values $G_1(t)=F_1(t), G_6(t), F_6(t)$;  (\textbf{b}) the mean values $G_3(t),F_3(t), G_4(t), F_4(t)$. In (\textbf{c}), the asymptotic value $\overline{G}_6$ of $G_6(t)$ versus the parameter $p_{3,3}^{(g)}$. Initial conditions $G_1(0)=F_1(0)=0.5$ and other mean values are equal to zero. Model parameters: all $p_{\alpha,\beta}^{(g,f)}=0.5$ in Equations \eqref{l33}--\eqref{l36}, $p_{3,3}^{(g)}=2$, $p_{4,4}^{(f)}=0.05$, all $\omega^{(g)}_\alpha=\omega^{(f)}_\alpha=1$ in Equation \eqref{l32}.} 
 \label{fig:sim1}
 \end{figure}

 In this first experiment, we present some results derived from the mechanisms described above. The outcomes are shown in Figures \ref{fig:sim1}a,b, where  the time evolution of the relevant mean values is depicted for an initial condition where $G_1(0)=F_1(0)=0.5$, and the model parameters are chosen as described in the 
captions of the figures. Note that, with respect to Section~\ref{sec:model}, we focus our attention on the transmitter $\cal T$ and the receiver $\cal R$, the intermediate layer being less relevant. For this reason, the figures in this Section describe mostly what happens to these agents. 
 
We set the strength of transmission, governed by the parameters $p_{\alpha,\beta}^{(g,f)}$, to be equal for all agents except for Agents 3 and 4, which are responsible for changing the nature of the transmitted information. We choose $p_{3,3}^{(g)}=2$ and $p_{4,4}^{(f)}=0.05$, meaning that there is a greater probability that the information arriving at Agent 3 will be changed from fake to good and a lower probability that Agent 4 will change from good to fake. This explains why the final asymptotic value of $G_6$ is higher than that of $F_6$.
Notice from Figure \ref{fig:sim1} that $G_1(t)=F_1(t)$ because the first layer conveys information to the middle layer with equal strength for both good and fake information, i.e., $p_{\alpha,1}^{(g)}=p_{\alpha,1}^{(f)}$ for $\alpha=2,\ldots,5$. In Figure \ref{fig:sim1}b, we observe that $G_{3,4}(t)$ and $F_{3,4}(t)$ initially increase due to the transmission from the first layer and then start to decrease after reaching their peaks, which is also true for the other mean values of the middle layer (not shown in the figure). It is $G_3(t)>F_3(t)$ because of the higher value of $p_{3,3}^{(g)}$, while $F_4(t)\gtrapprox G_4(t)$ due to the small value of $p_{4,4}^{(f)}$.

It is clear from this experiment that the ratio between $p_{3,3}^{(g)}$ and $p_{4,4}^{(f)}$ plays a crucial role in determining the final asymptotic value $\overline{G}_6$ of $G_6(t)$ (the asymptotic value of $F_6(t)$ in this case is simply $1-\overline{G}_6$). In Figure \ref{fig:sim1}c, we show $\overline{G}_6$ versus $p_{3,3}^{(g)}$ for the same initial conditions and model parameters, as reported in the caption. As $p_{3,3}^{(g)}$ increases, so does $\overline{G}_6$, although we observe a saturation effect for very large values of $p_{3,3}^{(g)}$, with a value close to $0.624$. Clearly, modifying the other parameters and initial conditions will affect this value. 

 \subsection{Experiment II}
\begin{figure}[H] 
\centering
 {\includegraphics[width=0.49\textwidth]{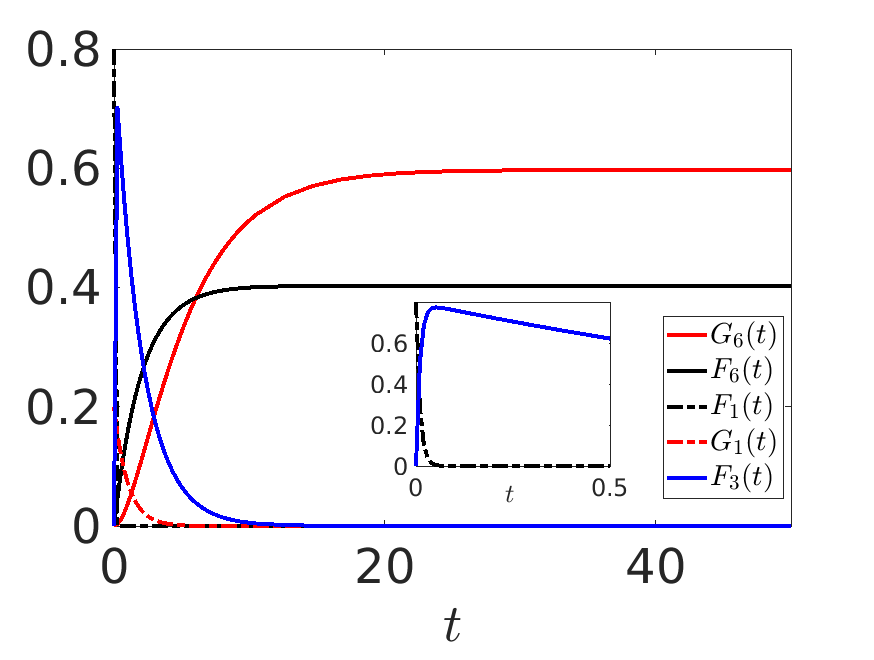}}
 \caption{Time evolutions of the mean values $G_1(t),F_1(t), G_6(t), F_6(t)$ and $F_3(t)$. Model parameters: all $p_{\alpha,\beta}^{(g,f)}=0.5$ in Equations \eqref{l33}--\eqref{l36}, except $p_{1,3}^{(f)}=5$, $p_{3,3}^{(g)}=2$, $p_{4,4}^{(f)}=0.05$; all other parameters as in Experiment I. In the inset, the time evolution for small times of $F_1(t)$ and $F_3(t)$.} \label{fig:sim3a}
 \end{figure}

 The following  numerical experiment is intended to show that, from an initial condition such that $G_1(0)<F_1(0)$, and choosing suitable strengths of the various Lindblad operators, it is possible to obtain 
 $G_6(t)>F_6(t)$ asymptotically, so that the nature of the transmitted information is viewed differently from the final receiver. This is similar to what is observed in Section \ref{sec:model}, for instance, adopting rule 1 or rule 2. Following the mechanism described in the previous Sections, we choose $p_{4,4}^{(f)}=0, p_{1,3}^{(f)}=5,$ and the other parameters as in Experiment I. Results are shown in Figure \ref{fig:sim3a} for the initial condition $G_1(0)=0.2$ and $F_1(0)=0.8$. With this choice of parameters, we are forcing the information to travel rapidly toward Agent $3$ due to the high $p_{1,3}^{(f)}=5$. Considering also $p_{3,3}^{(g)}=2>p_{4,4}^{(f)}$, this agent is forced to modify the nature of the information and transmit it modified as ``good''. The final receiver is then receiving more ``good'' information, expressed by the fact that $G_6(t)>F_6(t)$ asymptotically: the original nature of $\cal N$ is indeed modified by the interactions acting in $\cal S$. 
 
\subsection{Experiment III}
For our third experiment, we propose a variation of the model by introducing a new mechanism. While, in the previous experiments, all initial information in the first layer was entirely sent to the final receiver $\cal R$, this is not necessarily always the case. Sometimes, news or information is continuously transmitted for a longer period of time. To model this, we require the transmitter $\cal T$ to be continuously refilled with packets of information to send to the lower layers, and this can be easily implemented by using the Lindbladian approach. For example, an additional Lindblad operator of the form $L^{(g)}=\lambda_1^{(g)}g_1^\dag$ will push a pure state $\Psi=\sum\alpha_{\bf{n,m}}\Psi_{\mathbf{n,m}} $ with $\bf{ n}$ of the form $(0,\ldots)$ into the state $\Psi=\sum\alpha_{\bf{n,m}}\Psi_{\mathbf{n^\star,m}} $ with $\bf{ n}^\star$ of the form $(1,\ldots)$ (similarly, $L^{(f)}=\lambda_1^{(f)}f_1^\dag$ will push a pure state $\Psi=\sum\alpha_{\bf{m,n}}\Psi_{\mathbf{m,n}} $ toward $\Psi=\sum\alpha_{\bf{m,n}}\Psi_{\mathbf{m,n^\star}} $, with obvious notation). The effect of this operator is that the value of $G_1(t)$, or $F_1(t)$,  increases over time, at least asymptotically. In other words, the introduction of these Lindblad operators allows for the continuous refurnishing of information packets to the transmitter $\cal T$, which in turn is expected to result in a gradual increase in the quantity of information transmitted to the lower layers of the network. Of course, the precise dynamics of this process will depend on various factors, such as the strength of the Lindblad operator and the initial conditions of the system.

\begin{figure}[H] 
\centering
{\captionsetup{position=bottom,justification=centering}
 \subfloat[]{\includegraphics[width=0.49\textwidth]{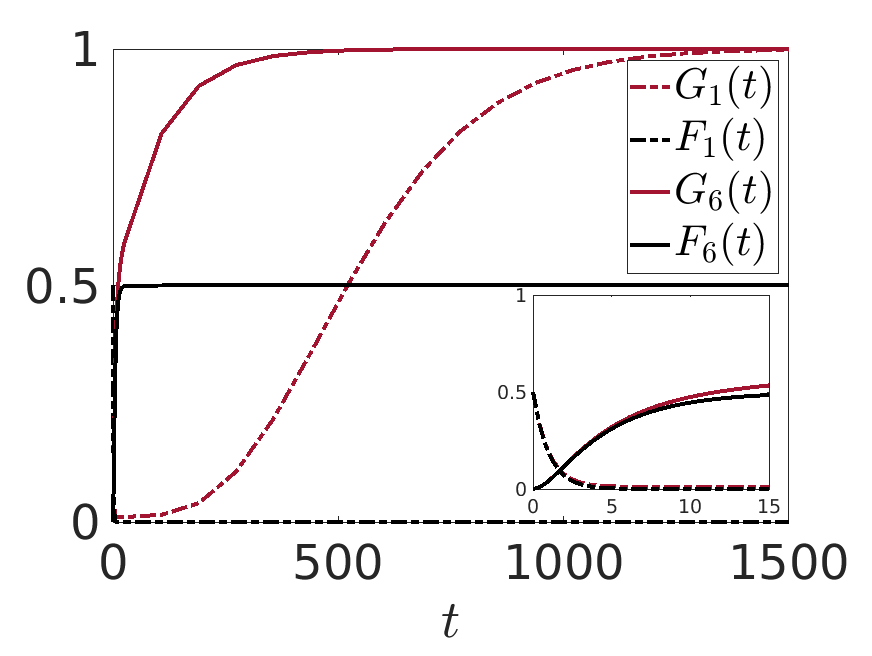}}
 \subfloat[]{\includegraphics[width=0.49\textwidth]{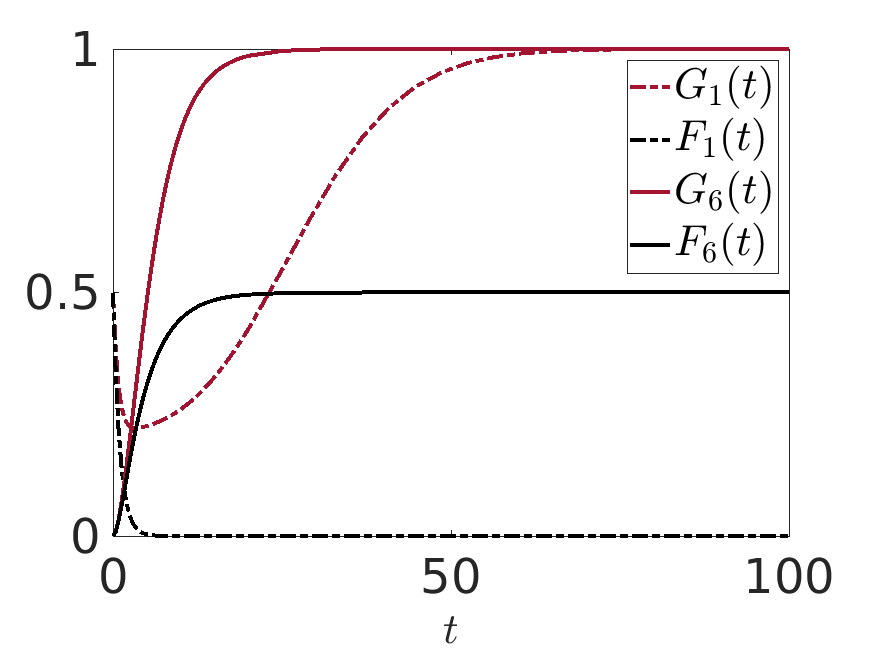}}}
\caption{(\textbf{a}) Time evolutions of the mean values $G_1(t),F_1(t), G_6(t), F_6(t)$ when considering the Lindblad operators $\lambda_1^{(g)}g_1^\dagger$. Model parameters: $\lambda_1^{(g)}=0.1$, $p_{3,3}^{(g)}=0$, $p_{4,4}^{(f)}=0.0$; all other parameters as in Experiment I. In the inset, the time evolution for small time (\textbf{b}). Same as panel (\textbf{a}) but with $\lambda_1^{(g)}=0.5$.} \label{fig:sim3b}
\end{figure}

Some numerical results are shown in Figures \ref{fig:sim3b}a,b, where we have chosen 
$\lambda_1^{(g)}\neq0, \lambda_1^{(f)}=0$. Our focus is solely on the pure transmission of the information, without any alteration of its inherent nature, and, to ensure this, we set the values of $p_{3,3}^{(g)}$ and $p_{4,4}^{(f)}$ to be equal to zero. All the other model parameters and initial conditions are consistent with those used in Experiment I.
As expected,  the mean value $G_1(t)$, after a small transient, increases and reaches the maximum value $1$. This has as a consequence that the lower layers continuously receive the ``good'' information, so that $G_6(t)$ reaches asymptotically the maximum value $1$ also. Figure \ref{fig:sim3b}a,b also show that the larger the value of $\lambda_1^{(g)}$, the higher the speed with which this asymptotic value is reached.

\section{Conclusions}
\label{section:Concl}
In this paper, we have proposed two different operatorial approaches to derive the dynamics governing the transmission of information in a multi-layered network.

In the first approach, the dynamics is governed by the Heisenberg equations of motion for operators, coupled with the application of  some rules modifying the inertia parameters. The reliability, or lack thereof, of
information is determined through the computation of suitable mean values of certain number operators. The dynamics is highly enriched by the rules; several numerical experiments are presented to discuss the influence of these rules on the spreading of information. Notably, we observe that the rules may allow for asymptotic equilibria in the dynamics (a significant feature that could not be achieved using standard Heisenberg dynamics) in the present context.

The second approach introduced is based on  the GKLS equations, giving an alternative method to reach equilibrium in the dynamics. This method is mainly based on the definition of particular Lindblad operators and the ideas typically characterizing the open quantum systems. The central idea here is that Lindblad operators enable the irreversibility of the transmission of information from an initial transmitter to a final receiver. This feature relaxes the constraint of the self-adjointness of the Hamiltonian present in the  $(H,\rho)$-induced dynamics approach, creating a direct way to describe the mechanisms of the transmission of the information within the network.

Both approaches allow for the derivation of plausible dynamical behaviors in which the receiver of the information can perceive it differently, depending on the choice of the parameters in the model.
Our analysis paves the way for various potential applications, ranging from the implementation of different rules to the modeling of specialized agents like influencers within the network. Moreover, our models are adaptable to other systems, and we see a robust connection to game theory as a promising future direction.
These explorations are central to our upcoming plans and represent only a small fraction of the possible extensions of the concepts discussed in this paper, and they underscore the versatility and potential of both the $(H,\rho)$-induced dynamics and the GKLS equations in the complex study of networked systems.

{The two strategies that we present are rather different, and also the results are qualitatively different, at least during the transient. Nonetheless, both are able to produce a time evolution approaching some equilibrium, which is an interesting feature that is impossible to achieve for finite-dimensional systems driven by self-adjoint Hamiltonians. Deeper investigations are required to obtain a valuable comparison between the two strategies.

A final comment is concerned with the observables that we are mainly interested in, in the description of the status of the various agents in the network, e.g., the mean value of the number of operators associated with fake and good news. Other quantities could be considered---for instance, entropies. Since a piece of news can assume two different states, using the mean values of fake and good news as probabilities, we could consider, for instance, Shannon entropy, and investigate its time evolution for single agents and for the network as a whole. We plan to investigate such an aspect in the near future. }

\section*{Acknowledgments}
F.B. and F.G. acknowledge partial financial support from Palermo University (via FFR2023 ``Bagarello'' and FFR2023 ``Gargano''). M.G. and F.O.  acknowledge 
partial support from the University of Messina.
M.G. acknowledges financial support from \textit{{Finanziamento} del Programma Operativo Nazionale (PON) ``Ricerca e Innovazione'' 2014-2020 a valere sull’Asse IV ``Istruzione e ricerca per il recupero''---Azione IV---Dottorati e contratti di ricerca su tematiche dell'innovazione, CUP J45F21001750007}.
All authors acknowledge partial financial support from G.N.F.M. of the INdAM. The work of all authors has been partially supported by the PRIN  grant \textit{``{Transport} phenomena in low dimensional structures: models, simulations and theoretical aspects''}.


\begin{thebibliography}{99}
	
\bibitem{N1} Jin %MDPI: Comment for Author Self-Citation
%We find that in the references, there are xxx references from you or where you are a co-author (Refs. 5, 6, 8, 12, 13).
%According to the journal’s citation policy and COPE guidelines (https://www.mdpi.com/journal/sustainability/instructions), we suggest that authors reduce cite their own references. Could you please replace some of these references with other relevant sources? Thank you very much for your understanding and cooperation in advance.
% Answer: Refs. 5, 6, 8, 12, 13 are essential for a reader to better understand the techniques used and to make comparisons between the approaches already used in literature.
, F.; Dougherty, E.; Saraf, P.; Cao, Y.; Ramakrishnan, N. {Epidemiological %MDPI: Please DO NOT change/revert the form of references in Reference Section, they have been completed layout and ready for publication. Please just provide the detailed information if required in the comments below. Or please provide the website links and accessed date (Day Month Year) if you cannot provide detailed information. We have done layout work for this manuscript and remove the link. Please modify it directly in the file. Please NEVER use EndNote or other tools to rearrange the reference order. Please revise ref in this version. 1. References in the main text are cited in numerical order; 2. Reference cited in the main text match the reference in the Reference List.
 Modeling of News and Rumors on Twitter}, In Proceedings of the SNAKDD '13: Proceedings of the 7th Workshop on Social Network Mining and Analysis, Chicago, IL, USA, 11~August~{2013}; Volume 8, pp. 1--8.%MDPI: We added the location and date of the conference. Please confirm.
% Answer: We confirm that.

\bibitem{N2} Lerman, K. {Social Information Processing in News Aggregation}. \emph{IEEE Internet Comput.} \textbf{2007}, \emph{11}, 16--28.

\bibitem{N3} Abdullah, S.; Wu, X. {An Epidemic Model for News Spreading on Twitter}. In Proceedings of the IEEE 23rd International Conference on Tools with Artificial Intelligence, Boca Raton, FL, USA, 7--9 November 2001; pp. 163--169.
%MDPI: We added the location and date of the conference. Please confirm.
% Answer: We confirm that.

\bibitem{N4} Doerr, B.; Fouz, M.; Friedrich, T. {Why rumors spread so quickly in social networks}. \emph{Commun. ACM} \textbf{2012}, \emph{55}, 70--75.%MDPI: Please add page number.

% Answer: We added the page number.
		
\bibitem{fff1}  Bagarello, F.; Gargano, F.; Oliveri, F. {Spreading of competing information on a network}. \emph{Entropy} \textbf{2020}, {\emph 22}, 1169.
	
\bibitem{bookBGO} Bagarello, F.; Gargano, G.; Oliveri, F. \emph{ Quantum Tools for Macroscopic Systems}; Synthesis Lectures on Mathematics \& Statistics; Springer: Cham, Switzerland, 2023. %MDPI: We added the location of the publisher. Please confirm.
% Answer: We confirm that.
	
\bibitem{Rob2023} Robinson, T.R. \emph{The Quantum Nature of Things: How Counting Leads to the Quantum}; CRC Press: Boca Raton, FL, USA, 2023.

\bibitem{HRO2} Bagarello, F.; Salvo, R.D.; Gargano, F.; Oliveri, F. {$(H,\rho)$--induced dynamics and large time behaviors}. \emph{Phys. A} {\bf 2018}, \emph{505}, 355--373.
	
\bibitem{Manz}Manzano, D. { A Short Introduction to the Lindblad Master Equation}. \emph{AIP Adv.} \textbf{2020}, \emph{10}, 025106.


\bibitem{bagbook1} Bagarello, F. \emph{Quantum Dynamics for Classical
Systems: With Applications of the Number Operator}; Wiley: New York, NY, USA, 2012.
	
\bibitem{bagbook2} Bagarello, F. \emph{Quantum Concepts in the Social, Ecological and Biological Sciences}; Cambridge University Press: Cambridge, UK, 2019.%MDPI: We added the location of the publisher. Please confirm.
% Answer: We confirm that.
	
\bibitem{DSGO2020} Di Salvo, R.; Gorgone, M.; Oliveri, F. {Generalized Hamiltonian for a two-mode fermionic model and asymptotic equilibria}. \emph{Phys. A} {\bf 2020}, \emph{540}, 12032.
		
\bibitem{BaGa23} Bagarello, F.; Gargano, F. {Dynamics for a quantum parliament}. \emph{Stud. Appl. Math.} \textbf{2023}, \emph{150}, 1182--1200.%MDPI: Please add volume number, or doi information.
% We added volume and inserted the corrected page number.

\bibitem{khren1} Asano, M.; Ohya, M.; Tanaka, Y.; Basieva, I.; Khrennikov, A. {Quantum-like model of brain's functioning: Decision making from decoherence}. \emph{J. Theor. Biol.} {\bf 2011}, \emph{281}, 56--64.


\bibitem{Asan2013} Asano, M.; Basieva, I.; Ohya, A.K.; Tanaka, Y.; Yamato, I. {A model of epigenetic evolution based on theory of open quantum systems}. \emph{Syst. Synth. Biol.} \textbf{2013}, \emph{7}, 161--173.

\bibitem{Nava2022} Nava, A.; Giuliano, D.; Papa, A.; Rossi, M. {Traffic models and traffic-jam transition in quantum (N+1)-level systems}. \emph{SciPost Phys. Core} \textbf{2022}, \emph{5}, 22.

\bibitem{Nava2023} Nava, A.; Giuliano, D.; Papa, A.; Rossi, M. {Understanding traffic jams using lindblad superoperators}. \emph{Int. J. Theor. Phys.} \textbf{2023}, \emph{62},~2.

\bibitem{Krea23} Basieva, I.; Khrennikov, A. {``What Is Life?'': Open Quantum Systems Approach}. \emph{Open Syst. Inf. Dyn.} \textbf{2023}, \emph{29}, 2250016.		

\end{thebibliography}
\end{document}